\documentclass[journal]{IEEEtran}

\usepackage{amsmath,amssymb,bm,mathtools}
\usepackage{amsthm}
\usepackage{graphicx}
\usepackage{grffile}

\usepackage{cite}

\theoremstyle{definition}
\newtheorem{theorem}{Theorem}[section]
\newtheorem{lemma}{Lemma}[section]
\newtheorem{example}{Example}[section]
\newtheorem{assumption}{Assumption}[section]
\newtheorem{claim}{Claim}[section]

\usepackage{algpseudocode,algorithm}

\newcommand{\minimize}{\operatornamewithlimits{\mathrm{minimize}}}
\newcommand{\argmin}{\operatornamewithlimits{\mathrm{arg\ min}}}
\newcommand{\plim}{\operatornamewithlimits{\mathrm{plim}}}
\newcommand{\calS}{\mathcal{S}}

\newcommand{\calT}{\mathcal{T}}
\newcommand{\sign}{\mathrm{sign}}
\newcommand{\subto}{\mathrm{subject\ to\ }}
\newcommand{\txc}{\text{c}}
\newcommand{\txe}{\text{e}}
\newcommand{\txu}{\text{u}}
\newcommand{\txv}{\text{v}}
\newcommand{\prox}{\mathrm{prox}}
\newcommand{\Pto}{\xrightarrow{\mathrm{P}}}
\newcommand{\abs}[1]{\left\lvert#1\right\rvert}
\newcommand{\norm}[1]{\left\lVert#1\right\rVert}
\newcommand{\Ex}[1]{\mathrm{E}\left[#1\right]}
\newcommand{\paren}[1]{\left(#1\right)}
\newcommand{\sqbra}[1]{\left[#1\right]}
\newcommand{\curbra}[1]{\left\{#1\right\}}

\allowdisplaybreaks[1]

\begin{document}
\title{Asymptotic Performance Prediction\\for ADMM-Based Compressed Sensing}
%
\author{
	Ryo~Hayakawa,~\IEEEmembership{Member,~IEEE}
	\thanks{
        This work was supported in part by JSPS KAKENHI Grant Number JP20K23324.%
	}%
	\thanks{
		Ryo Hayakawa is with Graduate School of Engineering Science, Osaka University, Osaka 560-8531, Japan (e-mail: hayakawa.ryo.es@osaka-u.ac.jp). 
	}
    \thanks{
        \copyright\ 2022 IEEE. 
        Personal use of this material is permitted. 
        Permission from IEEE must be obtained for all other uses, in any current or future media, including reprinting/republishing this material for advertising or promotional purposes, creating new collective works, for resale or redistribution to servers or lists, or reuse of any copyrighted component of this work in other works.
    }
}%
\markboth{}%
{R. Hayakawa: Asymptotic Performance Prediction for ADMM-Based Compressed Sensing}
%
\maketitle
\begin{abstract}
In this paper, we propose a method to predict the asymptotic performance of the alternating direction method of multipliers (ADMM) for compressed sensing, where we reconstruct an unknown structured signal from its underdetermined linear measurements. 
The derivation of the proposed method is based on the recently developed convex Gaussian min-max theorem (CGMT), which can be applied to various convex optimization problems to obtain its asymptotic error performance. 
Our main idea is to analyze the convex subproblem in the update of ADMM iteratively and characterize the asymptotic distribution of the tentative estimate obtained at each iteration. 
However, since the original CGMT cannot be used directly for the analysis of the iterative updates, we intuitively assume an extended version of CGMT in the derivation of the proposed method. 
Under the assumption, the result shows that the update equations in ADMM can be decoupled into a scalar-valued stochastic process in the asymptotic regime with the large system limit. 
From the asymptotic result, we can predict the evolution of the error (e.g., mean-square-error (MSE) and symbol error rate (SER)) in ADMM for large-scale compressed sensing problems. 
Simulation results show that the empirical performance of ADMM and its prediction are close to each other in sparse vector reconstruction and binary vector reconstruction. 
\end{abstract}
%
\IEEEpeerreviewmaketitle
%
\section{Introduction} \label{sec:intro}
\IEEEPARstart{C}{ompressed} sensing~\cite{candes2005,candes2006,donoho2006,candes2008} becomes a key technology in the field of signal processing such as image processing~\cite{lustig2007,lustig2008} and wireless communication~\cite{hayashi2013,choi2017}. 
A basic problem in compressed sensing is to reconstruct an unknown sparse vector from its underdetermined linear measurements, where the number of measurements is less than that of unknown variables. 
The compressed sensing techniques take advantage of the sparsity as the prior knowledge to reconstruct the vector. 
The idea of compressed sensing can also be applied for other structured signals by appropriately utilizing the structures, e.g., group sparsity~\cite{huang2010}, low-rankness~\cite{candes2009,candes2010}, and discreteness~\cite{aissa-el-bey2015,nagahara2015}. 

For compressed sensing, various algorithms have been proposed in the literature. 
Greedy algorithms such as matching pursuit (MP)~\cite{mallat1993} and orthogonal matching pursuit (OMP)~\cite{pati1993,tropp2007} iteratively update the support of the estimate of the unknown sparse vector. 
Several improved greedy algorithms have also been proposed to achieve better reconstruction performance~\cite{dai2009,needell2009,liu2012,donoho2012,wang2012,kwon2014}. 
Another approach for compressed sensing is based on message passing algorithms using a Bayesian framework. 
Approximated belief propagation (BP)~\cite{kabashima2003} and approximate message passing (AMP)~\cite{donoho2009} can reconstruct the unknown vector with low computational complexity. 
Moreover, the asymptotic performance can be predicted by the state evolution framework~\cite{donoho2009,bayati2011}. 
The AMP algorithm can also be used when the unknown vector has some structure other than the sparsity~\cite{jeon2015,hayakawa2018}. 
However, the AMP algorithm requires an assumption on the measurement matrix, and hence other message passing-based algorithms have also been proposed~\cite{cespedes2014,rangan2017,ma2017,takeuchi2020a}. 

Convex optimization-based approaches have also been well studied for compressed sensing. 
The most popular convex optimization problem for compressed sensing is the $\ell_{1}$ optimization, where we utilize the $\ell_{1}$ norm as the regularizer to promote the sparsity of the estimate. 
Although the objective function is not differentiable, the iterative shrinkage thresholding algorithm (ISTA)~\cite{daubechies2004,combettes2005,figueiredo2007} and the fast iterative shrinkage thresholding algorithm (FISTA)~\cite{beck2009} can solve the $\ell_{1}$ optimization problem with feasible computational complexity. 
Another promising algorithm is the alternating direction method of multipliers (ADMM)~\cite{gabay1976,eckstein1992,combettes2011,boyd2011}, which can be applied to a wider class of optimization problems than ISTA and FISTA. 
Moreover, ADMM can provide a sufficiently accurate solution with a relatively small number of iterations in practice~\cite{boyd2011}. 
However, since the convergence speed largely depends on the parameter, it is important to determine the appropriate parameter in practical applications. 
For the parameter selection of ADMM, several approaches have been proposed~\cite{raghunathan2014,ghadimi2015,xu2017,xu2017b,lin2017}. 
However, they are rather heuristic or inapplicable to compressed sensing problems. 

There are several theoretical analyses for convex optimization-based compressed sensing, e.g.,~\cite{donoho2011,bayati2012,donoho2013}. 
In particular, recently developed convex Gaussian min-max theorem (CGMT)~\cite{thrampoulidis2015d,thrampoulidis2018} can be utilized to obtain the asymptotic error of various optimization problems in a precise manner. 
In the CGMT-based analysis, we firstly derive an auxiliary optimization problem for the original optimization problem to be analyzed. 
We then investigate the properties of the optimizer of the auxiliary optimization. 
By using CGMT, we can relate the optimizer of the auxiliary optimization and that of the original optimization to obtain some analytical results of the original optimization problem (For details, see Appendix~\ref{app:CGMT}).
For example, the asymptotic mean-square-error (MSE) has been analyzed for various regularized estimators~\cite{atitallah2017,thrampoulidis2018}. 
The asymptotic symbol error rate (SER) has also been derived for convex optimization-based discrete-valued vector reconstruction~\cite{thrampoulidis2018a,hayakawa2020a}.  
The CGMT-based analysis has been extended for the optimization problem in the complex-valued domain~\cite{abbasi2019}, whereas the above analyses consider optimization problems in the real-valued domain. 
These analyses focus on the performance of the optimizer and do not deeply discuss the optimization algorithm to obtain the optimizer. 
The performance analysis for the optimization algorithms would be useful for algorithm choice, parameter tuning, development of a new algorithm, and so on. 

In this paper, as an attempt to analyze optimization algorithms, we propose a method to predict the asymptotic behavior of ADMM for convex optimization-based compressed sensing. 
The main idea is that, when we use the squared loss function as the data fidelity term, the subproblem in the iterations of ADMM has a similar form to the optimization problems that can be analyzed by the CGMT framework. 
Based on the fact that the tentative estimate is the optimizer of the subproblem, we aim to analyze the subproblem to track the behavior of the estimate in ADMM. 
However, since the original CGMT cannot be applied directly to the subproblem in an iterative manner, we intuitively assume an extended version of CGMT in this paper. 
In the derivation under the assumption, we firstly derive the auxiliary optimization problem for the subproblem. 
We then analyze the auxiliary optimization problem and utilize the assumption of the extended CGMT to apply the analytical result to the optimizer of the subproblem. 
As a result, we claim that the asymptotic distribution of the tentative estimate can be characterized by a scalar-valued stochastic process, which depends on the measurement ratio, the parameter in the optimization problem, the parameter in ADMM, the distribution of the unknown vector, and the noise variance. 
As a corollary, we also predict the evolution of the error such as MSE and SER in ADMM for large-scale compressed sensing problems. 
We can utilize the asymptotic result to reveal the effect of the parameter in ADMM and tune it to achieve fast convergence. 

As examples, we consider sparse vector reconstruction and binary vector reconstruction and then evaluate the asymptotic result via computer simulations. 
Simulation results show that the asymptotic evolution of MSE converges to the MSE of the optimizer, which can be obtained with the previous CGMT-based analysis in the literature~\cite{thrampoulidis2018}. 
We also observe that the empirical performance of ADMM and its prediction are close to each other in both sparse vector reconstruction and binary vector reconstruction, though the assumption of the extended CGMT has not been proven rigorously. 

The rest of the paper is organized as follows. 
In Section~\ref{sec:ADMM}, we describe the ADMM-based compressed sensing as preliminary. 
We then provide the main results for ADMM in Section~\ref{sec:result} and the proposed performance prediction method in Section~\ref{sec:proposed}. 
In Section~\ref{sec:examples}, we consider two examples of the reconstruction problem and show several simulation results. 
Finally, Section~\ref{sec:conclusion} presents some conclusions. 

In this paper, we use the following notations. 
We denote the transpose by $(\cdot)^{\top}$ and the identity matrix by $\bm{I}$. 
For a vector $\bm{z}=\sqbra{z_{1}\, \cdots \, z_{N}}^{\top} \in \mathbb{R}^{N}$, the $\ell_{1}$ norm and the $\ell_{2}$ norm are given by $\norm{\bm{z}}_{1} = \sum_{n=1}^{N} \abs{z_{n}}$ and $\norm{\bm{z}}_{2} = \sqrt{\sum_{n=1}^{N} z_{n}^{2}}$, respectively. 
We denote the number of nonzero elements of $\bm{z}$ by $\norm{\bm{z}}_{0}$. 
$\sign(\cdot)$ denotes the sign function. 
For a lower semicontinuous convex function $\zeta: \mathbb{R}^{N} \to \mathbb{R} \cup \curbra{+\infty}$, we define the proximity operator as $\prox_{\zeta} (\bm{z}) = \argmin_{\bm{u} \in \mathbb{R}^{N}} \curbra{ \zeta(\bm{u}) + \frac{1}{2} \norm{\bm{u}-\bm{z}}_{2}^{2}}$. 
The Gaussian distribution with mean $\mu$ and variance $\sigma^{2}$ is denoted as $\mathcal{N}(\mu, \sigma^{2})$. 
When a sequence of random variables $\curbra{\Theta_{n}}$ ($n=1,2,\dotsc$) converges in probability to $\Theta$, we denote $\Theta_{n} \Pto \Theta$  as $n \to \infty$ or $\plim_{n \to \infty} \Theta_{n} = \Theta$. 
%
\section{ADMM-Based Compressed Sensing} \label{sec:ADMM}
In this paper, we consider the reconstruction of an $N$ dimensional vector $\bm{x} = \sqbra{x_{1}\ \dotsb\ x_{N} }^{\top} \in \mathbb{R}^{N}$ from its linear measurements given by 
\begin{align}
	\bm{y} = \bm{A} \bm{x} + \bm{v} \in \mathbb{R}^{M}. 
\end{align}
Here, $\bm{A} \in \mathbb{R}^{M \times N}$ is a known measurement matrix and $\bm{v} \in \mathbb{R}^{M}$ is an additive Gaussian noise vector. 
We denote the measurement ratio by $\Delta = M / N$. 
In the scenario of compressed sensing, we focus on the underdetermined case with $\Delta < 1$ and utilize the structure of $\bm{x}$ as the prior knowledge for the reconstruction. 
Note that we can use not only the sparsity but also other structures such as boundedness and discreteness~\cite{tan2001, nagahara2015}. 

Convex optimization is a promising approach for compressed sensing because we can flexibly design the objective function to utilize the structure of the unknown vector $\bm{x}$. 
In this paper, we consider the following convex optimization problem 
\begin{align}
    \minimize_{\bm{s} \in \mathbb{R}^{N}} 
    \curbra{
        \frac{1}{2} \norm{ \bm{y}-\bm{A}\bm{s} }_{2}^{2} 
        + \lambda f ({\bm{s}}) 
    }, \label{eq:optimization}
\end{align}
where $f(\cdot): \mathbb{R}^{N} \to \mathbb{R} \cup \curbra{+\infty}$ is a convex regularizer to utilize the prior knowledge of the unknown vector $\bm{x}$. 
For example, $\ell_{1}$ regularization $f(\bm{s}) = \norm{\bm{s}}_{1}$ is a popular convex regularizer for the reconstruction of the sparse vector. 
The regularization parameter $\lambda$ ($>0$) controls the balance between the data fidelity term $\dfrac{1}{2} \norm{\bm{y} - \bm{A} \bm{s}}_{2}^{2}$ and the regularization term $\lambda f({\bm{s}})$. 

ADMM has been used in a wide range of applications because it can be applied to various optimization problems. 
Moreover, we can obtain a sufficiently accurate solution with a relatively small number of iterations in practice~\cite{boyd2011}. 
To derive ADMM for the optimization problem~\eqref{eq:optimization}, we firstly rewrite~\eqref{eq:optimization} as 
\begin{align}
    &\minimize_{\bm{s}, \bm{z} \in \mathbb{R}^{N}} 
    \curbra{
        \frac{1}{2} \norm{\bm{y} - \bm{A} \bm{s}}_{2}^{2} 
        + \lambda f ({\bm{z}}) 
    } \notag \\
    &\ \subto 
    \ \bm{s} = \bm{z}. \label{eq:optimization_ADMM}
\end{align}
The update equations of ADMM for~\eqref{eq:optimization_ADMM} are given by 
\begin{align}
    \bm{s}^{(k+1)} 
    &= 
    \argmin_{\bm{s} \in \mathbb{R}^{N}} 
    \curbra{
        \frac{1}{2} \norm{\bm{y} - \bm{A} \bm{s}}_{2}^{2} + \frac{\rho}{2} \norm{\bm{s} - \bm{z}^{(k)} + \bm{w}^{(k)}}_{2}^{2} 
    } \label{eq:ADMM_update_s1} \\
    &= 
    \paren{\bm{A}^{\top} \bm{A} + \rho \bm{I}}^{-1} 
    \paren{\bm{A}^{\top} \bm{y} + \rho \paren{\bm{z}^{(k)} - \bm{w}^{(k)}}}, \label{eq:ADMM_update_s2} \\
    \bm{z}^{(k+1)} 
    &= 
    \argmin_{\bm{z} \in \mathbb{R}^{N}} 
    \curbra{
        \lambda f(\bm{z}) + \frac{\rho}{2} \norm{\bm{s}^{(k+1)} - \bm{z} + \bm{w}^{(k)}}_{2}^{2} 
    } \label{eq:ADMM_update_z} \\
    &= 
    \prox_{\frac{\lambda}{\rho} f} \paren{\bm{s}^{(k+1)} + \bm{w}^{(k)}}, \label{eq:ADMM_update_z2} \\
    \bm{w}^{(k+1)} 
    &= 
    \bm{w}^{(k)} + \bm{s}^{(k+1)} - \bm{z}^{(k+1)}, \label{eq:ADMM_update_w}
\end{align}
where $k$ ($= 0, 1, 2, \dotsc$) is the iteration index in the algorithm and $\rho$ ($>0$) is the parameter. 
In this paper, we refer to $\bm{s}^{(k+1)}$ as the tentative estimate of the unknown vector $\bm{x}$ in ADMM. 
We use $\bm{z}^{(0)} = \bm{w}^{(0)} = \bm{0}$ as the initial value in this paper. 
%
\section{Main Result} \label{sec:result}
The main purpose of this paper is the prediction of the error performance of ADMM for compressed sensing. 
In this section, we provide the main result for the behavior of ADMM in~\eqref{eq:ADMM_update_s1}--\eqref{eq:ADMM_update_w} for the problem~\eqref{eq:optimization_ADMM}. 

\subsection{Assumptions}
In the analysis, we use the following assumptions. 
\begin{assumption} \label{ass:distribution}
    The unknown vector $\bm{x}$ is composed of independent and identically distributed (i.i.d.) random variables with a known distribution $p_{X}$ which has some mean and variance. 
    The measurement matrix $\bm{A} \in \mathbb{R}^{M \times N}$ is composed of i.i.d.\ Gaussian random variables with zero mean and variance $1/N$. 
    Moreover, the additive noise vector $\bm{v} \in \mathbb{R}^{M}$ is also Gaussian with zero mean and the covariance matrix $\sigma_{\txv}^{2}\bm{I}$. 
    The unknown vector $\bm{x}$, the measurement matrix $\bm{A}$, and the additive noise vector $\bm{v}$ are independent. 
\end{assumption}
\begin{assumption} \label{ass:regularizer}
    The regularizer $f(\cdot): \mathbb{R}^{N} \to \mathbb{R} \cup \curbra{+\infty}$ is a lower semicontinuous convex function. 
    Moreover, $f(\cdot)$ is separable and expressed with the corresponding function $\tilde{f}(\cdot): \mathbb{R} \to \mathbb{R} \cup \curbra{+\infty}$ as $f(\bm{s}) = \sum_{n=1}^{N} \tilde{f}(s_{n})$, where $\bm{s} = \sqbra{s_{1}\ \dotsb\ s_{N}}^{\top} \in \mathbb{R}^{N}$. 
    With the slight abuse of notation, we use the same $f(\cdot)$ for the corresponding function $\tilde{f}(\cdot)$. 
\end{assumption}

In Assumption~\ref{ass:distribution}, we assume that the elements of the measurement matrix $\bm{A}$ are Gaussian variables because CGMT requires the Gaussian assumption in the proof~\cite{thrampoulidis2018}. 
However, the universality~\cite{bayati2015,panahi2017,oymak2018} of random matrices suggests that the result of the analysis can be applied when the measurement matrix is drawn from some other distributions. 
In fact, our result is valid for the random matrix from Bernoulli distribution with $\{1/\sqrt{N}, -1/\sqrt{N}\}$ in computer simulations (See Example~\ref{ex:sparse}). 

In Assumption~\ref{ass:regularizer}, we assume the separability of the regularizer $f(\cdot)$.  
Under this assumption, the proximity operator $\prox_{\gamma f}(\cdot): \mathbb{R}^{N} \to \mathbb{R}^{N}$ ($\gamma > 0$) becomes an element-wise function, i.e., the $n$-th element of the output depends only on the corresponding $n$-th element of the input. 

\subsection{Claim for Asymptotic Performance of ADMM for Compressed Sensing}
Under Assumptions~\ref{ass:distribution} and~\ref{ass:regularizer}, we consider the asymptotic behavior of ADMM. 
As in usual high-dimensional analyses~\cite{bayati2011,thrampoulidis2018}, we consider the sequence of problems with $\curbra{\bm{x}, \bm{A}, \bm{v}}$ indexed by $N$ and take the large system limit, where $N$ and $M$ go to infinity with a fixed ratio $\Delta=M/N$. 
In this paper, we denote the large system limit by $N \to \infty$ for simplicity. 

On the asymptotic behavior of ADMM, we provide the following claim. 
\begin{claim} \label{cl:main}
    We assume that $\bm{x}$, $\bm{A}$, $\bm{v}$, and $f(\cdot)$ satisfy Assumptions~\ref{ass:distribution} and~\ref{ass:regularizer}. 
    We consider the following stochastic process 
    \begin{align}
        S_{k+1} 
        &= 
        \hat{S}_{k+1} (\alpha_{k}^{\ast}, \beta_{k}^{\ast}), \label{eq:update_S} \\
        Z_{k+1} 
        &= 
        \prox_{\frac{\lambda}{\rho} f} \paren{ S_{k+1} + W_{k} }, \label{eq:update_Z} \\
        W_{k+1} 
        &= 
        W_{k} + S_{k+1} - Z_{k+1} \label{eq:update_W}
    \end{align}
    with the index $k$, where $\hat{S}_{k+1}(\alpha, \beta)$ is the function of $\alpha$ and $\beta$ defined as 
    \begin{align}
        &\hat{S}_{k+1} (\alpha, \beta) \notag \\
        &= 
        \dfrac{1}{\dfrac{\beta\sqrt{\Delta}}{\alpha} + \rho} 
        \paren{ 
            \dfrac{\beta\sqrt{\Delta}}{\alpha} 
            \paren{X + \dfrac{\alpha}{\sqrt{\Delta}} H}
            + \rho (Z_{k} - W_{k}) 
        } \label{eq:S_alpha_beta}
    \end{align}
    with the random variables $X \sim p_{X}$ and $H \sim \mathcal{N}(0, 1)$ ($Z_{0} = W_{0} = 0$). 
    We here assume the optimization problem 
    \begin{align}
        &\min_{\alpha>0} \max_{\beta \ge 0} 
        \curbra{
            \frac{\alpha\beta\sqrt{\Delta}}{2} 
            + \frac{\beta \sigma_{\txv}^{2} \sqrt{\Delta}}{2\alpha} 
            - \frac{1}{2} \beta^{2} 
            + \Ex{J^{(k+1)}(\alpha,\beta)} 
        } \label{eq:SO}
    \end{align}
    has a unique optimizer $(\alpha_{k}^{\ast}, \beta_{k}^{\ast})$, where 
    \begin{align}
        &J^{(k+1)} (\alpha, \beta) \notag \\
        &= 
        \frac{\beta\sqrt{\Delta}}{2\alpha} \paren{\hat{S}_{k+1}(\alpha, \beta) - X}^{2} 
        - \beta H \paren{\hat{S}_{k+1}(\alpha, \beta) - X} \notag \\
        &\hspace{4mm}+ 
        \frac{\rho}{2} \paren{ \hat{S}_{k+1}(\alpha, \beta) - Z_{k} + W_{k}}^{2} \label{eq:J}
    \end{align}
    is the function of $\alpha$ and $\beta$ to be optimized. 
    The expectation is taken over all random variables $X$, $H$, $Z_{k}$, and $W_{k}$. 
    
    Let $\mu_{\bm{s}^{(k+1)}}$ be the empirical distribution of $\bm{s}^{(k+1)} = \sqbra{s_{1}^{(k+1)}\ \dotsb\ s_{N}^{(k+1)}}^{\top}$ corresponding to the cumulative distribution function (CDF) given by $P_{\bm{s}^{(k+1)}}(s) = \frac{1}{N} \sum_{n=1}^{N} \mathbb{I} \paren{s_{n}^{(k+1)} < s}$, where we define $\mathbb{I}\paren{s_{n}^{(k+1)} < s} = 1$ if $s_{n}^{(k+1)} < s$ and otherwise $\mathbb{I}\paren{s_{n}^{(k+1)} < s} = 0$. 
    Moreover, we denote the distribution of the random variable $S_{k+1}$ in~\eqref{eq:update_S} as $\mu_{S_{k+1}}$. 
    Then, the empirical distribution $\mu_{\bm{s}^{(k+1)}}$ converges weakly in probability to $\mu_{S_{k+1}}$ as $N \to \infty$, i.e., $\int g d \mu_{\bm{s}^{(k+1)}} \Pto \int g d \mu_{S_{k+1}}$ holds for any continuous compactly supported function $g(\cdot): \mathbb{R} \to \mathbb{R}$. 
\end{claim}

For the detail of the derivation of the claim, see Appendix~\ref{app:derivation}. 
Although the idea of the procedure is based on the analysis via CGMT~\cite{thrampoulidis2018}, the original CGMT-based analysis cannot be applied directly to the case considered in this paper. 
We thus additionally assume an intuitively extended version of CGMT to derive the result of Claim~\ref{cl:main}. 
Therefore, it should be noted that the claim above has not been proven rigorously in part, though the simulation results in Section~\ref{sec:examples} demonstrate its effectiveness. 

\begin{figure*}[!t]
    \centering
    \includegraphics[width=160mm]{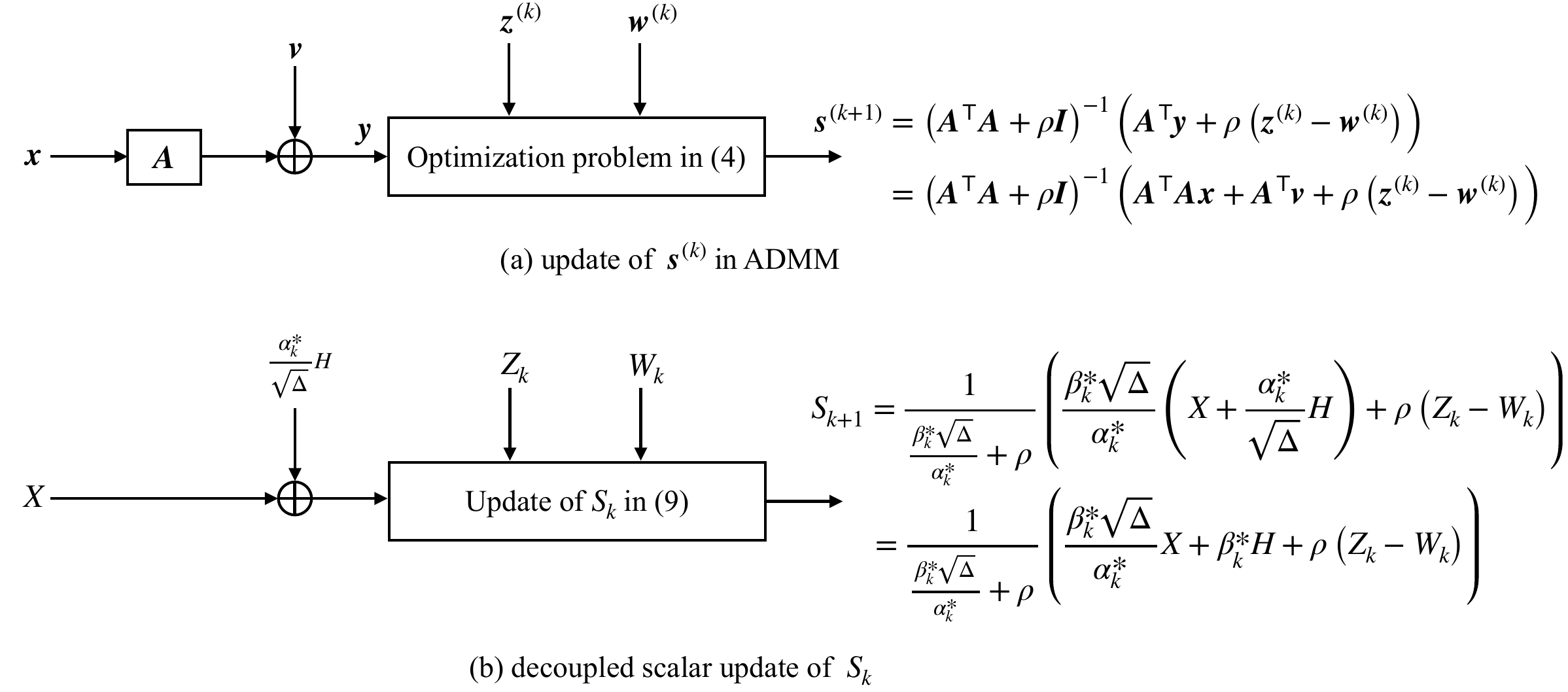}
    \caption{Comparison between the update of $\bm{s}^{(k)}$ and its decoupled version.}
    \label{fig:comparison_update}
\end{figure*}

Claim~\ref{cl:main} means that the distribution of the elements of $\bm{s}^{(k)}$ is characterized by the random variable $S_{k}$ in the asymptotic regime with $M, N \to \infty$ ($M /N = \Delta$). 
The update of $\bm{s}^{(k)}$ in~\eqref{eq:ADMM_update_s1} and~\eqref{eq:ADMM_update_s2} denotes the computation in the finite dimensional space, whereas the stochastic process in Claim~\ref{cl:main} gives the asymptotic and probabilistic property of the tentative estimate in ADMM. 
From the claim, we can see that the behavior of $S_{k}$ depends on the measurement ratio $\Delta$, the distribution of the unknown vector $p_{X}$, the noise variance $\sigma_{\txv}^{2}$, the regularization parameter $\lambda$, and the parameter of ADMM $\rho$. 
Moreover, the update of $S_{k}$ in~\eqref{eq:update_S} can be regarded as the `decoupled' version of the update of $\bm{s}^{(k)}$ in~\eqref{eq:ADMM_update_s1}, where the update is not element-wise. 
Figure~\ref{fig:comparison_update} shows the comparison between the update of $\bm{s}^{(k)}$ in~\eqref{eq:ADMM_update_s1} and its decoupled version obtained from Claim~\ref{cl:main}. 
In the update of $\bm{s}^{(k)}$, the measurement vector $\bm{y}$ is obtained through the linear transformation of $\bm{x}$ and additive Gaussian noise channel. 
On the other hand, in the decoupled system, the random variable $X$ goes through only the additive Gaussian noise channel. 
The variance of the additive Gaussian noise is $\paren{\alpha_{k}^{\ast}}^{2} / \Delta$, which implies that the noise variance in the decoupled system decreases as the measurement ratio $\Delta$ increases\footnote{Although the value of $\alpha_{k}^{\ast}$ depends on $\Delta$, $\alpha_{k}^{\ast}$ relates to the MSE as in Claim~\ref{cl:MSE}. 
Since the MSE decreases as the measurement ratio increases in general, $\alpha_{k}^{\ast}$ would also decrease in that case. 
}. 
We can also see that the update of $\bm{s}^{(k)}$ and $S_{k}$ have the similar form because they can be rewritten as
\begin{align}
    &\bm{s}^{(k+1)} \notag \\
    &= 
    \paren{\bm{A}^{\top} \bm{A} + \rho \bm{I}}^{-1} 
    \paren{\bm{A}^{\top} \bm{A} \bm{x} + \bm{A}^{\top} \bm{v} + \rho \paren{\bm{z}^{(k)} - \bm{w}^{(k)}}}, \\
    &S_{k+1} 
    = 
    \dfrac{1}{\dfrac{\beta_{k}^{\ast} \sqrt{\Delta}}{\alpha_{k}^{\ast}} + \rho} 
    \paren{ 
        \dfrac{\beta_{k}^{\ast} \sqrt{\Delta}}{\alpha_{k}^{\ast}} X 
        + \beta_{k}^{\ast} H 
        + \rho (Z_{k} - W_{k}) 
    }, 
\end{align}
respectively, despite the complicated derivation via the CGMT framework. 
The update of $S_{k}$ in~\eqref{eq:update_S} and~\eqref{eq:S_alpha_beta} shows that $S_{k+1}$ is the weighted sum of $X + \dfrac{\alpha_{k}^{\ast}}{\sqrt{\Delta}} H$ and $Z_{k} - W_{k}$ with the weights $\dfrac{\beta_{k}^{\ast} \sqrt{\Delta}}{\alpha_{k}^{\ast}}$ and $\rho$, respectively. 
Since $\rho$ is the parameter of ADMM, we can control the weight in the update of $S_{k}$ by tuning $\rho$. 
Similar decoupling phenomena also appear in other CGMT-based analyses (e.g.,~\cite{hayakawa2020a}) and AMP-based analyses (e.g.,~\cite{jeon2015}). 

One of the most important performance measures for the reconstruction algorithm is MSE given by $\dfrac{1}{N} \norm{\bm{s}^{(k)} - \bm{x}}_{2}^{2}$. 
As in the CGMT-based analysis~\cite{thrampoulidis2018}, the optimal value of $\alpha$ is related to the asymptotic MSE. 
Specifically, from Claim~\ref{cl:main}, we claim that the asymptotic MSE of the tentative estimate $\bm{s}^{(k)}$ in ADMM can be written with $\alpha_{k}^{\ast}$, which is the optimizer of~\eqref{eq:SO} (See also~\cite[Remark IV. 1]{hayakawa2020a})
\footnote{
Intuitively, from~\eqref{eq:root_eliminate_trick}, the optimal value $\hat{\alpha}$ of $\alpha$ in~\eqref{eq:root_eliminate_trick} satisfies $\hat{\alpha}^{2} = \norm{\bm{e}}_{2}^{2}/N + \sigma_{\txv}^{2}$, where $\bm{e}$ denotes the error $\bm{e} = \bm{s} -\bm{x}$. 
This formula implies that the square of the optimal $\alpha$ is the sum of the MSE and the noise variance, i.e., the MSE is the square of the optimal $\alpha$ minus the noise variance. 
}. 
Since Claim~\ref{cl:MSE} is based on Claim~\ref{cl:main}, it is also a partly unproven claim. 
\begin{claim} \label{cl:MSE}
    Under the assumptions in Claim~\ref{cl:main}, the asymptotic MSE of $\bm{s}^{(k+1)}$ is given by 
    \begin{align}
        \plim_{N \to \infty} 
        \frac{1}{N} \norm{\bm{s}^{(k+1)} - \bm{x}}_{2}^{2} 
        &= 
        \paren{\alpha_{k}^{\ast}}^{2} - \sigma_{\txv}^{2}. 
    \end{align}
\end{claim}
%
\section{Proposed Performance Prediction Method} \label{sec:proposed}
\subsection{Performance Prediction Based on Claim~\ref{cl:main}}
From the results of Claim~\ref{cl:main} or Claim~\ref{cl:MSE}, we can predict the performance of ADMM for large-scale compressed sensing problems. 
However, since the exact computation of the expectation in~\eqref{eq:SO} is difficult, we use its approximation alternatively to predict the performance as shown in Algorithm~\ref{alg:prediction}. 
\begin{algorithm}[t]
	\caption{Performance prediction based on Claim~\ref{cl:main}}
	\begin{algorithmic}[1]
		\Require{measurement ratio $\Delta$, distribution of unknown vector $p_{X}$, noise variance $\sigma_{\txv}^{2}$, regularization parameter $\lambda$, parameter of ADMM $\rho$, sample size $N_{\text{sample}}$}
		\State{Make $N_{\text{sample}}$ realizations of $X$ as $x^{[1]}, \dotsc, x^{[N_{\text{sample}}]}$ from the distribution $p_{X}$.}
		\State{Make $N_{\text{sample}}$ realizations of $H$ as $h^{[1]}, \dotsc, h^{[N_{\text{sample}}]}$ from the standard Gaussian distribution.}
		\State{Make $N_{\text{sample}}$ realizations of $Z_{0}$ and $W_{0}$ as $z_{0}^{[i]} = w_{0}^{[i]} = 0$ ($i=1, \dotsc, N_{\text{sample}}$).}
		\For{$k = 0, 1, \dotsc$}
		    \State{Solve~\eqref{eq:SO_prac} and obtain the optimal values $\hat{\alpha}_{k}^{\ast}$ and $\hat{\beta}_{k}^{\ast}$.}
		    \State{$s_{k+1}^{[i]} = \hat{s}_{k+1} \paren{\hat{\alpha}_{k}^{\ast}, \hat{\beta}_{k}^{\ast}, x^{[i]}, z_{k}^{[i]}, w_{k}^{[i]}, h^{[i]}}$\\ \hspace{17mm}($i=1, \dotsc, N_{\text{sample}}$)}
		    \State{$z_{k+1}^{[i]} = \prox_{\frac{\lambda}{\rho} f} \paren{s_{k+1}^{[i]} + w_{k}^{[i]}}$ ($i=1, \dotsc, N_{\text{sample}}$)}
		    \State{$w_{k+1}^{[i]} = w_{k}^{[i]} + s_{k+1}^{[i]} - z_{k+1}^{[i]}$ ($i=1, \dotsc, N_{\text{sample}}$)}
		    \State{Compute the the performance prediction with the realizations $s_{k+1}^{[i]}$ and $x^{[i]}$ ($i=1, \dotsc, N_{\text{sample}}$). }
		\EndFor
	\end{algorithmic}
	\label{alg:prediction}
\end{algorithm}
In the algorithm, after we make $N_{\text{sample}}$ realizations of the random variables $X$, $H$, $Z_{0}$, and $W_{0}$, we obtain the realizations of $S_{1}$, $Z_{1}$, $W_{1}$, $S_{2}$, $Z_{2}$, $W_{2}$, $\dotsc$ iteratively. 
At the $k$-th iteration, we solve 
\begin{align}
    &\min_{\alpha>0} \max_{\beta\ge0} 
    \Biggl\{
        \frac{\alpha\beta\sqrt{\Delta}}{2} 
        + \frac{\beta \sigma_{\txv}^{2} \sqrt{\Delta}}{2\alpha} 
        - \frac{1}{2} \beta^{2} \notag \\
    &\hspace{8mm}
        + \frac{1}{N_{\text{sample}}} \sum_{i = 1}^{N_{\text{sample}}} j^{(k+1)} \paren{\alpha, \beta, x^{[i]}, z_{k}^{[i]}, w_{k}^{[i]}, h^{[i]}} \label{eq:SO_prac}
    \Biggr\}
\end{align}
instead of~\eqref{eq:SO}, where we define 
\begin{align}
    &j^{(k+1)} (\alpha, \beta, x, z, w, h) \notag \\
    &= 
    \frac{\beta\sqrt{\Delta}}{2\alpha} \paren{\hat{s}_{k+1} (\alpha, \beta, x, z, w, h) - x}^{2} \notag \\
    &\hspace{4mm}-
    \beta h \paren{\hat{s}_{k+1} (\alpha, \beta, x, z, w, h) - x} \notag \\
    &\hspace{4mm}+ 
    \frac{\rho}{2} \paren{ \hat{s}_{k+1} (\alpha, \beta, x, z, w, h) - z + w}^{2} 
\end{align}
and 
\begin{align}
    &\hat{s}_{k+1} (\alpha, \beta, x, z, w, h) \notag \\
    &= 
    \dfrac{1}{\dfrac{\beta\sqrt{\Delta}}{\alpha} + \rho} 
    \paren{ 
        \dfrac{\beta\sqrt{\Delta}}{\alpha} 
        \paren{x + \dfrac{\alpha}{\sqrt{\Delta}} h}
        + \rho (z - w) 
    } 
\end{align}
from~\eqref{eq:J} and~\eqref{eq:S_alpha_beta}, respectively. 
In~\eqref{eq:SO_prac}, $x^{[i]}$, $z_{k}^{[i]}$, $w_{k}^{[i]}$, and $h^{[i]}$ denote the $i$-th realizations of $X$, $Z_{k}$, $W_{k}$, and $H$, respectively ($i = 1, \dotsc, N_{\text{sample}}$). 
If $N_{\text{sample}}$ is sufficiently large, the optimal values $\hat{\alpha}_{k}^{\ast}$ and $\hat{\beta}_{k}^{\ast}$ for~\eqref{eq:SO_prac} are expected to be good approximations of $\alpha_{k}^{\ast}$ and $\beta_{k}^{\ast}$, respectively. 
After we solve~\eqref{eq:SO_prac} via some searching technique such as ternary search and golden-section search~\cite{luenberger2008}, the realizations of $S_{k+1}$, $Z_{k+1}$, $W_{k+1}$ are computed as 
\begin{align}
    s_{k+1}^{[i]} 
    &= 
    \hat{s}_{k+1} \paren{\hat{\alpha}_{k}^{\ast}, \hat{\beta}_{k}^{\ast}, x^{[i]}, z_{k}^{[i]}, w_{k}^{[i]}, h^{[i]}}, \\
    z_{k+1}^{[i]} 
    &= 
    \prox_{\frac{\lambda}{\rho} f} \paren{s_{k+1}^{[i]} + w_{k}^{[i]}}, \\
    w_{k+1}^{[i]} 
    &= 
    w_{k}^{[i]} + s_{k+1}^{[i]} - z_{k+1}^{[i]} 
\end{align}
from~\eqref{eq:update_S}--\eqref{eq:update_W} ($i = 1, \dotsc, N_{\text{sample}}$). 
From Claim~\ref{cl:MSE}, the asymptotic MSE at the $k$-th iteration can be predicted as $\paren{\hat{\alpha}_{k}^{\ast}}^{2} - \sigma_{\txv}^{2}$. 
Moreover, since $s_{k}^{[i]}$ and $x^{[i]}$ are the realizations of the elements of $\bm{s}^{(k)}$ and $\bm{x}$, respectively, we can also predict some other performance metrics such as SER from the realizations (For example, see Example~\ref{ex:binary}).  

\subsection{Parameter Tuning Based on Predictions}
From the result of Claim~\ref{cl:main} (or Claim~\ref{cl:MSE}), we can tune the parameter $\rho$ in ADMM to achieve fast convergence. 
The conventional parameter tuning methods~\cite{raghunathan2014,ghadimi2015,xu2017,xu2017b,lin2017} focus on the difference between the tentative estimate and the optimizer of the optimization problem. 
On the other hand, the parameter tuning based on Claim~\ref{cl:main} can take account of the error from the true unknown vector in the asymptotic regime. 
Since the effect of $\rho$ to $\alpha_{k}^{\ast}$ and $\beta_{k}^{\ast}$ is complicated, the explicit expression of the optimal $\rho$ is difficult to obtain. 
By numerical simulations, however, we can obtain an appropriate value of $\rho$ by predicting the asymptotic performance of ADMM for various $\rho$ and choosing the candidate that achieves the fastest convergence. 
For instance, see Fig.~\ref{fig:MSE_vs_iteration_rho_sparse} in Example~\ref{ex:sparse}. 
%
\section{Examples} \label{sec:examples}
In this section, we consider two examples of the reconstruction problem and compare the empirical performance of ADMM and its prediction obtained by Claim~\ref{cl:main}. 
\begin{example}[Sparse Vector Reconstruction] \label{ex:sparse}
    The $\ell_{1}$ optimization 
    \begin{align}
        \minimize_{\bm{s} \in \mathbb{R}^{N}} 
        \curbra{
            \frac{1}{2} \norm{ \bm{y}-\bm{A}\bm{s} }_{2}^{2} 
            + \lambda \norm{\bm{s}}_{1} 
        } \label{eq:L1_optimization}
    \end{align}
    with the $\ell_{1}$ norm is the most popular convex optimization problem for sparse vector reconstruction. 
    The $\ell_{1}$ regularization promotes the sparsity of the estimate of the unknown vector in the reconstruction. 
    We here assume that the distribution of each element of the unknown vector $\bm{x}$ is known and independent of each other. 
    In the simulations, the distribution is given by the Bernoulli-Gaussian distribution as 
    \begin{align}
        p_{X}(x)
        &= 
        p_{0} \delta_{0}(x) + (1 - p_{0}) p_{H}(x), 
    \end{align}
    where $p_{0} \in (0, 1)$, $\delta_{0}(\cdot)$ denotes the Dirac delta function, and $p_{H}(\cdot)$ is the probability density function of the standard Gaussian distribution\footnote{
        To compute the realization of $X$ in the simulations, we firstly obtain a sample $d \in [0, 1]$ from the uniform distribution on $[0, 1]$. 
        We then set $X = 0$ if $d < p_{0}$, and otherwise we let $X$ be a sample from the standard Gaussian distribution. 
    }. 
    When $p_{0}$ is large, the unknown vector becomes sparse. 
    The proximity operator of the $\ell_{1}$ norm is given by
    \begin{align}
        \sqbra{\prox_{\gamma \norm{\cdot}_{1}} (\bm{r})}_{n}
        &= 
        \sign (r_{n}) \max(\abs{r_{n}} - \gamma, 0), \label{eq:prox_L1}
    \end{align}
    where $\bm{r} = \sqbra{r_{1}\ \dotsb\ r_{N}}^{\top} \in \mathbb{R}^{N}$, $\gamma > 0$, and $\sqbra{\cdot}_{n}$ denotes the $n$-th element of the vector. 
    By using~\eqref{eq:prox_L1}, we can perform ADMM in~\eqref{eq:ADMM_update_s1}--\eqref{eq:ADMM_update_w} for the $\ell_{1}$ optimization~\eqref{eq:L1_optimization}. 
    Claim~\ref{cl:main} enables us to predict the asymptotic behavior of ADMM for the $\ell_{1}$ optimization. 
    
    We first compare the empirical performance of the sparse vector reconstruction and its prediction obtained from Claim~\ref{cl:main}. 
    Figure~\ref{fig:MSE_vs_iteration_N_sparse} shows that the MSE performance of the sparse vector reconstruction, where $\Delta = 0.9$, $p_{0} = 0.8$, and $\sigma_{\txv}^{2} = 0.001$. 
    \begin{figure}[t]
        \centering
        \includegraphics[width=85mm]{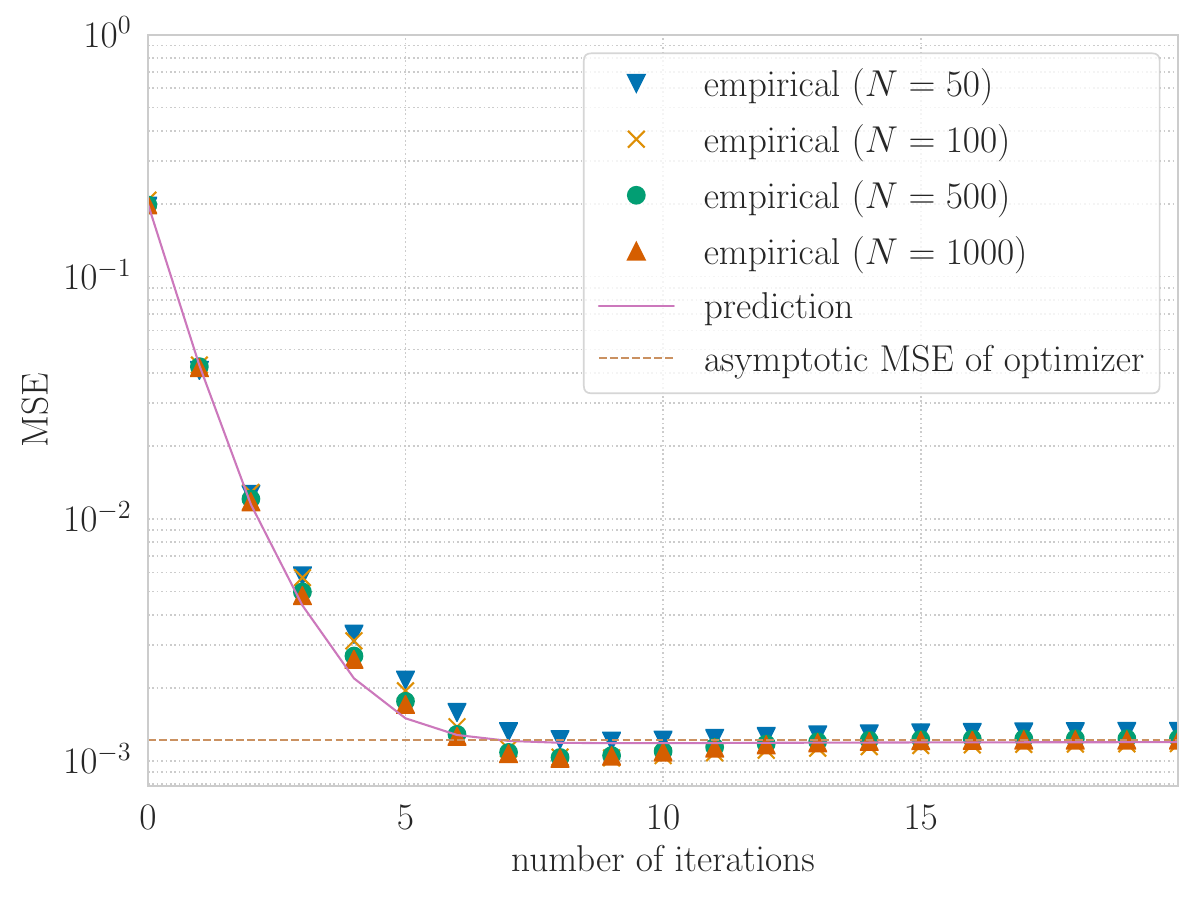}
        \caption{MSE performance for different $N$ in sparse vector reconstruction ($\Delta = 0.9$, $p_{0} = 0.8$, $\sigma_{\txv}^{2} = 0.001$, $\rho = 0.1$).}
        \label{fig:MSE_vs_iteration_N_sparse}
    \end{figure}
    The measurement matrix $\bm{A}$ and the noise vector $\bm{v}$ satisfy Assumption~\ref{ass:distribution}. 
    The parameter $\rho$ of ADMM is set as $\rho = 0.1$. 
    In the figure, `empirical' means the empirical performance obtained by ADMM in~\eqref{eq:ADMM_update_s1}--\eqref{eq:ADMM_update_w} when $N = 50, 100, 500$, and $1000$. 
    The empirical performance is obtained by averaging the results for $100$ independent realizations of $\bm{x}$, $\bm{A}$, and $\bm{v}$. 
    On the other hand, `prediction' shows the performance prediction obtained by Claim~\ref{cl:main} (or Claim~\ref{cl:MSE}) in the large system limit. 
    Note that the prediction is obtained independently of the empirical performance. 
    Since the exact computation of the distribution of $(S_{k}, Z_{k}, W_{k})$ is difficult in practice, we use Algorithm~\ref{alg:prediction} to make $N_{\text{sample}} = 100,000$ realizations of the random variables $(S_{k}, Z_{k}, W_{k})$ and obtain the approximation of $(\alpha_{k}^{\ast}, \beta_{k}^{\ast})$. 
    To optimize $\alpha$ and $\beta$, we can use searching techniques such as ternary search and golden-section search~\cite{luenberger2008}. 
    In the simulations, we use the ternary search with the error tolerance $10^{-6}$. 
    Although we have assumed that $\alpha_{k}^{\ast}$ and $\beta_{k}^{\ast}$ are uniquely determined in Claim~III.1, these optimizers are actually obtained in the simulations considered in this paper. 
    In Fig.~\ref{fig:MSE_vs_iteration_N_sparse}, we also show the asymptotic MSE of the optimizer obtained by applying CGMT to the $\ell_{1}$ optimization problem as in~\cite{thrampoulidis2018}. 
    It should be noted that the asymptotic MSE of the optimizer does not depend on the optimization algorithm. 
    The parameter $\lambda$ in~\eqref{eq:optimization} is determined by minimizing the asymptotic MSE. 
    From Fig.~\ref{fig:MSE_vs_iteration_N_sparse}, we can see that the empirical performance and its prediction are close to each other. 
    Moreover, they converge to the asymptotic MSE of the optimizer in the original $\ell_{1}$ optimization problem. 
    Precisely, there is a slight difference between the empirical performance and its prediction. 
    One possible reason is that the empirical performance is evaluated for finite $N$, whereas the large system limit $N \to \infty$ is assumed in the asymptotic analysis. 
    Another reason is that we create the many realizations of $(S_{k}, Z_{k}, W_{k})$ for the prediction instead of computing their exact distributions. 

    Next, we evaluate the MSE performance for different matrix structures. 
    Figure~\ref{fig:MSE_vs_iteration_matrix_sparse} shows the MSE performance when $N = 500$, $M = 250$, $p_{0} = 0.9$, $\sigma_{\txv}^{2} = 0.001$, and $\rho = 0.1$. 
    \begin{figure}[t]
        \centering
        \includegraphics[width=85mm]{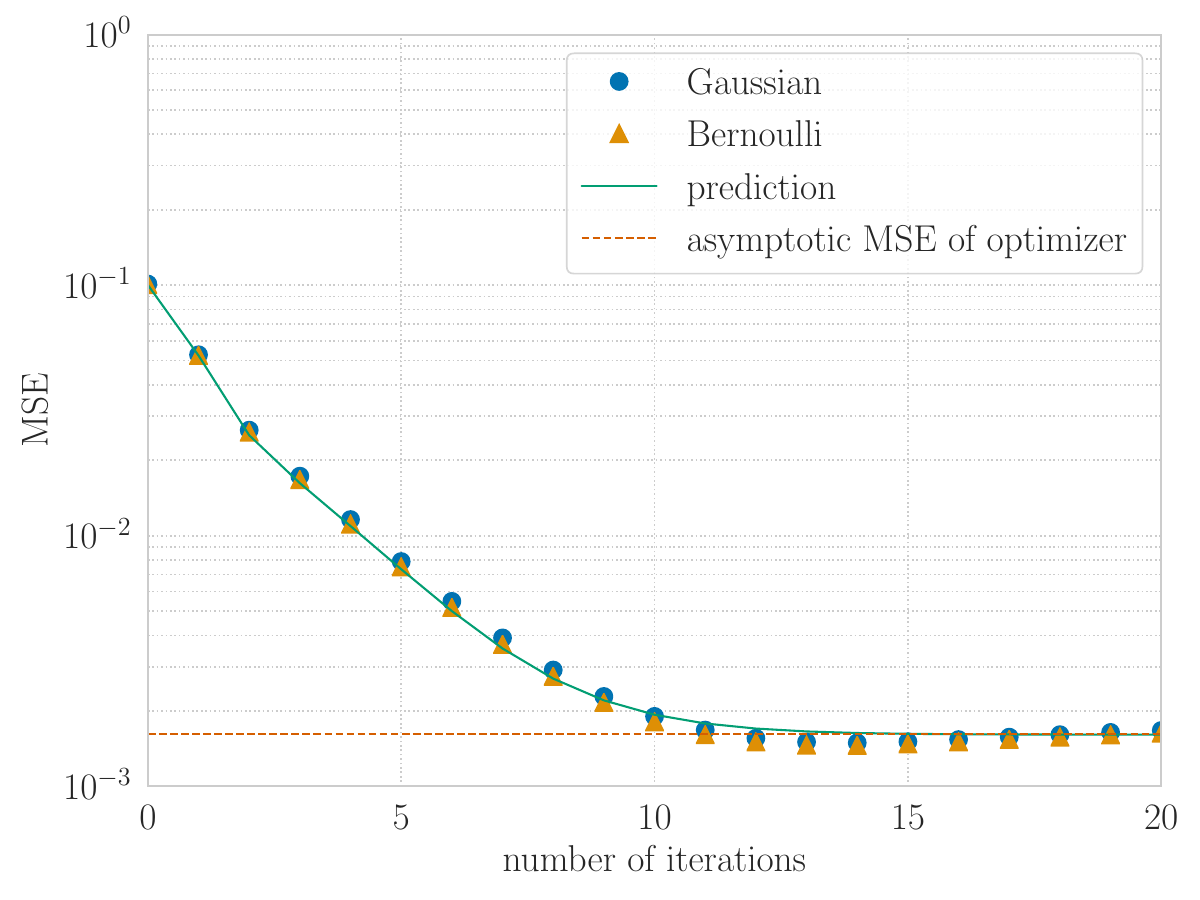}
        \caption{MSE performance for different measurement matrix in sparse vector reconstruction ($N = 500$, $M = 250$, $p_{0} = 0.9$, $\sigma_{\txv}^{2} = 0.001$, $\rho = 0.1$).}
        \label{fig:MSE_vs_iteration_matrix_sparse}
    \end{figure}
    In the figure, `Gaussian' means the performance when the measurement matrix $\bm{A}$ is composed of i.i.d.\ Gaussian elements and satisfies Assumption~\ref{ass:distribution}. 
    On the other hand, `Bernoulli' shows the performance when each element of the measurement matrix is drawn uniformly from $\{1/\sqrt{N}, -1/\sqrt{N}\}$. 
    The empirical performance is obtained by averaging the results for $500$ independent realizations of $\bm{x}$, $\bm{A}$, and $\bm{v}$. 
    From Fig.~\ref{fig:MSE_vs_iteration_matrix_sparse}, we observe that the empirical performance for both cases is close to the prediction obtained by Claim~\ref{cl:main} (or Claim~\ref{cl:MSE}). 
    
    We then evaluate the effects of the parameter $\rho$ in ADMM. 
    Figure~\ref{fig:MSE_vs_iteration_rho_sparse} shows the asymptotic MSE performance for $\rho=0.05, 0.2$, and $0.5$. 
    \begin{figure}[t]
        \centering
        \includegraphics[width=85mm]{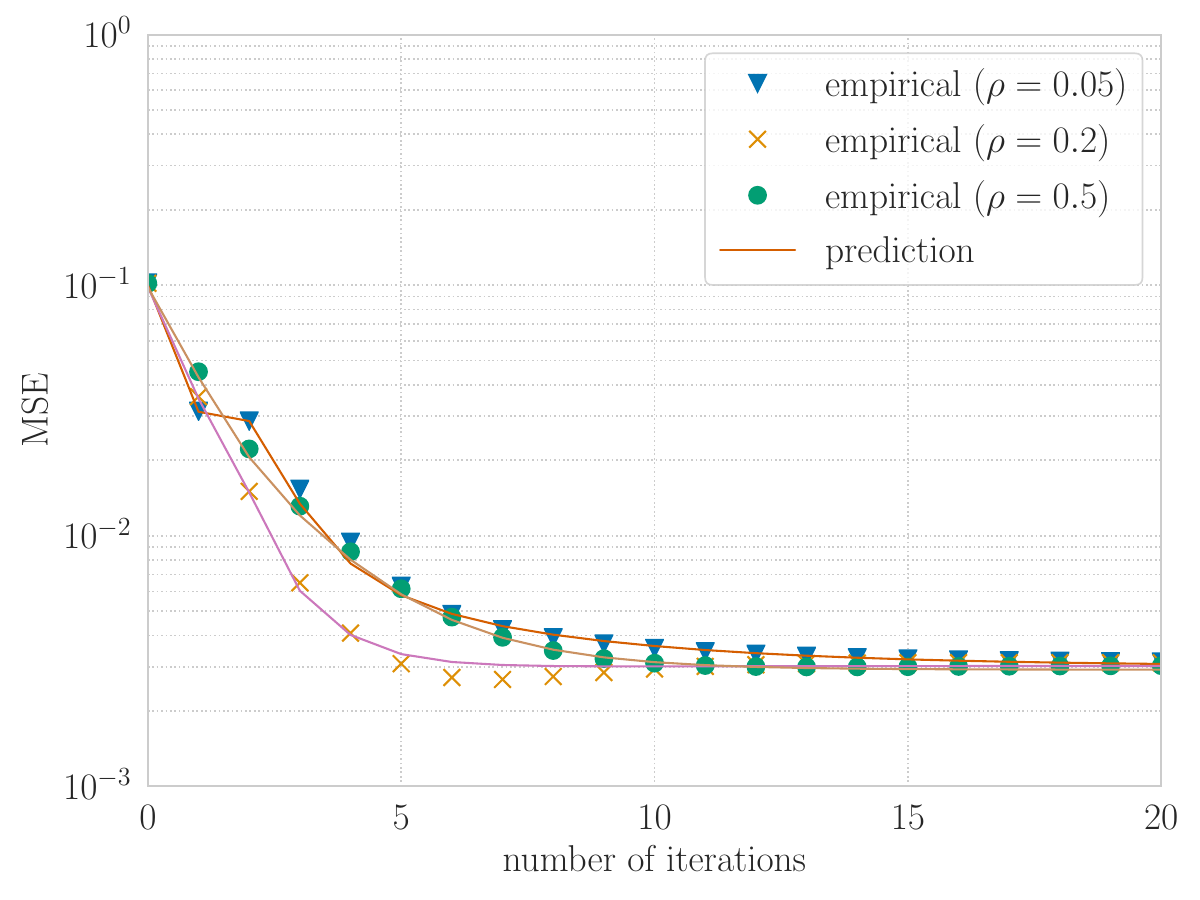}
        \caption{MSE performance for different parameter $\rho$ in sparse vector reconstruction ($N = 500$, $\Delta = 0.8$, $p_{0} = 0.9$, $\sigma_{\txv}^{2} = 0.005$).}
        \label{fig:MSE_vs_iteration_rho_sparse}
    \end{figure}
    In the figure, we set $N=500, \Delta = 0.8$, $p_{0} = 0.9$, and $\sigma_{\txv}^{2} = 0.005$. 
    We can see that the value of the parameter $\rho$ significantly affects the convergence speed of ADMM. 
    By using the prediction obtained from Claim~\ref{cl:main}, we can adjust $\rho$ to achieve fast convergence without the empirical reconstruction and the computation of MSE, which is unknown in practice. 
    
    Finally, we show the performance for various measurement ratios. 
    Figure~\ref{fig:MSE_vs_delta_sparse} shows the MSE performance versus $\Delta$ when $N = 500$, $p_{0} = 0.85$, $\sigma_{\txv}^{2} = 0.001$, and $\rho = 0.1$. 
    \begin{figure}[t]
        \centering
        \includegraphics[width=85mm]{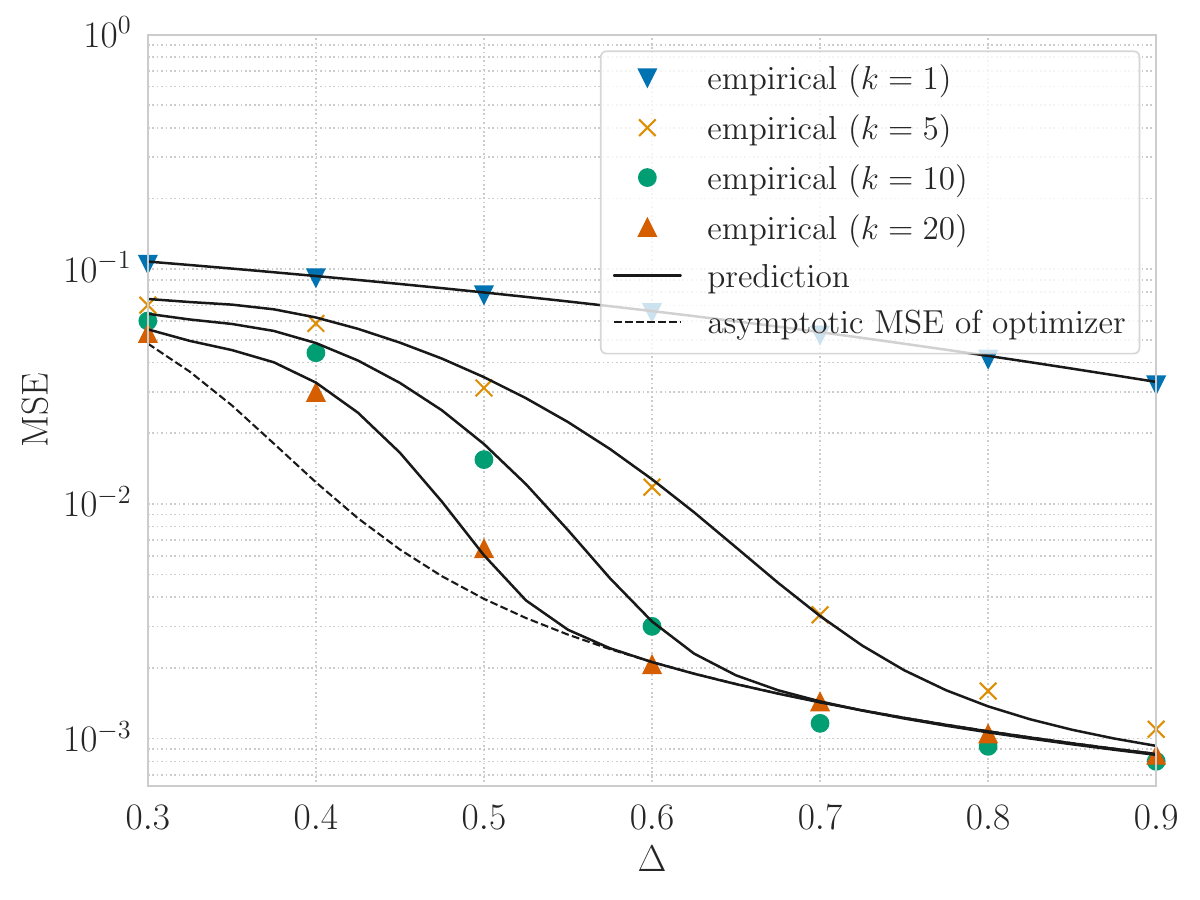}
        \caption{MSE performance versus $\Delta$ in sparse vector reconstruction ($N = 500$, $p_{0} = 0.85$, $\sigma_{\txv}^{2} = 0.001$, $\rho = 0.1$).}
        \label{fig:MSE_vs_delta_sparse}
    \end{figure}
    The empirical performance is evaluated by averaging the results for $500$ independent realizations of $\bm{x}$, $\bm{A}$, and $\bm{v}$, which satisfy Assumption~\ref{ass:distribution}. 
    The performance prediction is obtained in the same way as in Fig.~\ref{fig:MSE_vs_iteration_N_sparse}. 
    From the figure, we can see that the empirical MSE decreases as the iteration index $k$ increases for every $\Delta$. 
    Moreover, the empirical performance agrees well with the prediction at each iteration. 
    As expected, the MSE performance becomes better as the measurement ratio $\Delta$ increases. 
    The figure also shows that the ADMM algorithm converges already at the iteration index $k = 10$ for $\Delta \ge 0.7$, whereas it does not achieve the optimal MSE (dashed line) even at $k = 20$ for $\Delta =0.4, 0.5$. 
    This would be mainly because the parameter $\rho$ is fixed as $\rho = 0.1$ for simplicity. 
    If we choose an appropriate value of $\rho$ for each $\Delta$, the algorithm might achieve the optimal MSE at an earlier iteration even for such a small measurement ratio. 
\end{example}
\begin{example}[Binary Vector Reconstruction] \label{ex:binary}
    We consider the reconstruction of a binary vector $\bm{x} \in \{1, -1\}^{N}$ with the known distribution 
    \begin{align}
        p_{X}(x) 
        &= 
        \frac{1}{2} (\delta_{0}(x - 1) + \delta_{0}(x + 1)). 
    \end{align}
    A reasonable approach to reconstruct $\bm{x} \in \{1, -1\}^{N}$ is the box relaxation method~\cite{tan2001,yener2002} given by 
    \begin{align}
        \minimize_{\bm{s} \in \sqbra{-1, 1}^{N}} 
        \curbra{
            \frac{1}{2} \norm{ \bm{y}-\bm{A}\bm{s} }_{2}^{2} 
        }, \label{eq:box_optimization}
    \end{align}
    which is a convex relaxation of the maximum likelihood approach
    \begin{align}
        \minimize_{\bm{s} \in \curbra{1, -1}^{N}} 
        \curbra{
            \frac{1}{2} \norm{ \bm{y}-\bm{A}\bm{s} }_{2}^{2} 
        }. 
    \end{align}
    The asymptotic performance of the final estimate obtained by the box relaxation method has been analyzed with CGMT in~\cite{thrampoulidis2018a}. 
    The optimization problem~\eqref{eq:box_optimization} is equivalent to~\eqref{eq:optimization} with $f(\bm{s}) = \sum_{n=1}^{N} \iota(s_{n})$, where 
    \begin{align}
        \iota (s) 
        &= 
        \begin{cases}
            0 & (s \in \sqbra{-1, 1}) \\
            \infty & (s \notin \sqbra{-1, 1})
        \end{cases}. 
    \end{align}
    Since the proximity operator of $\iota(\cdot)$ is given by the projection to $\sqbra{-1, 1}$, i.e., 
    \begin{align}
        \prox_{\gamma \iota} (r) 
        &= 
        \min(\max(r, -1), 1), \label{eq:prox_box}
    \end{align}
    we can perform ADMM in~\eqref{eq:ADMM_update_s1}--\eqref{eq:ADMM_update_w} by using~\eqref{eq:prox_box}. 
    From Claim~\ref{cl:main}, we can predict the asymptotic performance of ADMM for the box relaxation method. 

    We evaluate the SER performance defined as $\norm{\sign\paren{\bm{s}^{(k)}} - \bm{x}}_{0} / N$, which is important performance measure in binary vector reconstruction. 
    In Algorithm~\ref{alg:prediction}, we can predict the asymptotic SER performance as $ \sum_{i=1}^{N_{\text{sample}}} \norm{\sign\paren{s_{k}^{[i]}} - x^{[i]}}_{0} / N_{\text{sample}}$. 
    Although the sign function is not continuous, we can approximate the function to use the result of Claim~\ref{cl:main} (cf. \cite[Lemma A.4]{thrampoulidis2018a}). 
    Figure~\ref{fig:SER_vs_iteration_delta_binary} shows the SER performance of ADMM for $\Delta = 0.7, 0.8$, and $0.9$, where $N = 500$, $\sigma_{\txv}^{2} = 0.04$, and $\rho = 0.1$. 
    \begin{figure}
        \centering
        \includegraphics[width=85mm]{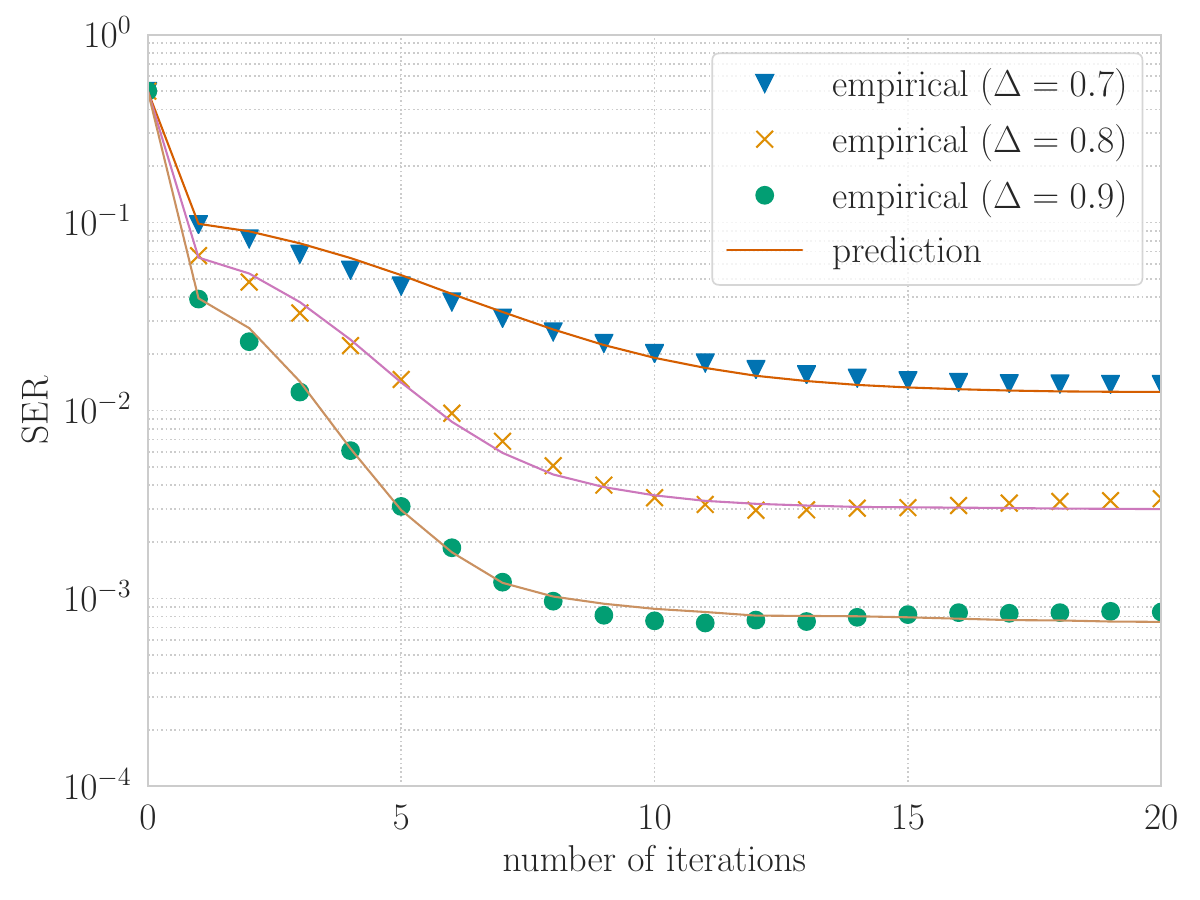}
        \caption{Asymptotic SER performance for different measurement ratio $\Delta$ in binary vector reconstruction ($N = 500$, $\sigma_{\txv}^{2} = 0.04$, $\rho = 0.1$).}
        \label{fig:SER_vs_iteration_delta_binary}
    \end{figure}
    The empirical performance is obtained by averaging $300$ results for independent realizations of $\bm{x}$, $\bm{A}$, and $\bm{v}$. 
    The prediction is computed by making $N_{\text{sample}} = 300,000$ realizations of the random variables $(S_{k}, Z_{k}, W_{k})$ via Algorithm~\ref{alg:prediction}. 
    We observe that the empirical performance and the prediction are close to each other. 
    We can see that the prediction of Claim~\ref{cl:main} is valid for the binary vector reconstruction.  
    
    Next, we compare the distributions of $\bm{s}^{(k)}$ in ADMM and $S_{k}$ in~\eqref{eq:update_S}. 
    Figure~\ref{fig:distribution_binary} shows the histogram of the empirical CDF $P_{\bm{s}^{(k)}}(s)$ and its prediction $P_{S_{k}}(s)$, where $N = 500$, $M = 400$, $\sigma_{\txv}^{2} = 0.001$, and $\rho = 0.1$. 
    \begin{figure*}
        \centering
        \includegraphics[width=180mm]{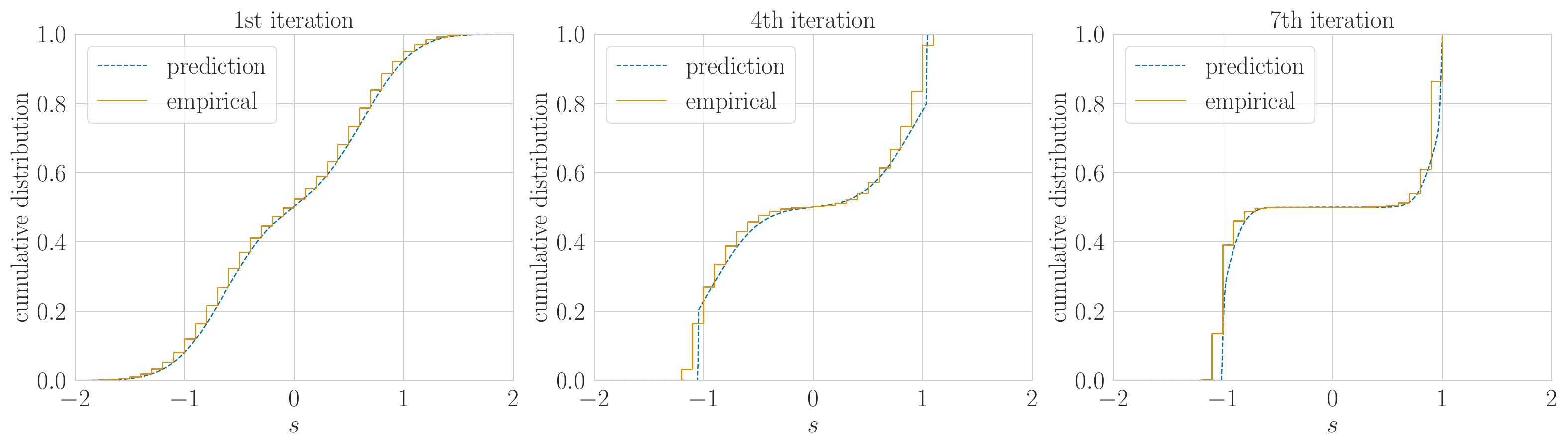}
        \caption{Comparison between empirical CDF and its prediction in binary vector reconstruction ($N = 500$, $M = 400$, $\sigma_{\txv}^{2} = 0.001$, $\rho = 0.1$).}
        \label{fig:distribution_binary}
    \end{figure*}
    The left, middle, and right figure denotes the distributions when $k=1$, $k=4$, and $k=7$, respectively. 
    The empirical performance is obtained by averaging $100$ results for independent realizations of $\bm{x}$, $\bm{A}$, and $\bm{v}$. 
    The prediction is computed by making $N_{\text{sample}} = 100,000$ realizations of the random variables $(S_{k}, Z_{k}, W_{k})$. 
    From Fig.~\ref{fig:distribution_binary}, we observe that the empirical CDF agrees well with the prediction at each iteration. 
    We can also see that the distributions concentrate near $1$ and $-1$ as the iteration proceeds. 
\end{example}
%
\section{Conclusion and Future Work} \label{sec:conclusion}
In this paper, we have proposed the prediction method for the asymptotic behavior of ADMM for compressed sensing. 
By using the recently developed CGMT framework, we have provided the stochastic process $\curbra{S_{k}}_{k=1,2,\dotsc}$ corresponding to the tentative estimate in ADMM. 
The main claim provides the prediction of the error evolution of ADMM in the large system limit. 
We can also tune the parameter in ADMM from the asymptotic result. 
Simulation results show that the empirical performance obtained by ADMM and its prediction are close to each other in terms of MSE and SER in sparse vector reconstruction and binary vector reconstruction, respectively. 

We here show some possible research directions based on the analysis in this paper. 
Although we consider the fixed parameter $\rho$ in ADMM, it is possible to use the different parameter $\rho_{k}$ at each iteration and predict the asymptotic performance in the same manner. 
The prediction in this case would provide the faster convergence of the algorithm. 
Moreover, both ADMM and CGMT can be applied to the convex optimization problem in the complex-valued domain~\cite{li2015,hayakawa2018b,abbasi2019}. 
It would also be an interesting topic to analyze the performance of ADMM for compressed sensing problems in the complex-valued domain, which often appear in communication systems. 

Further understanding of the relation between the two processes in Fig.~\ref{fig:comparison_update} would be important to clarify the behavior of ADMM for compressed sensing. 
A clear interpretation of the stochastic process in~\eqref{eq:update_S}--\eqref{eq:update_W} and the scalar optimization problem~\eqref{eq:SO} would also help us to understand the result in more detail. 

In the derivation of Claim~\ref{cl:main}, we have introduced the extended version of CGMT as an unproven assumption (For details, see Appendix~\ref{app:derivation}). 
The discussion on the necessity of the assumption and its proof (if possible) are also important for the theoretical validation of the results in this paper. 
\appendices
\section{CGMT} \label{app:CGMT}
In this section, we briefly summarize CGMT~\cite{thrampoulidis2015d,thrampoulidis2018,thrampoulidis2018a}, which has been proposed as a tool for the asymptotic analysis of the regularized convex optimization such as~\eqref{eq:optimization}. 
CGMT is based on the Gaussian min-max theory (GMT)~\cite{gordon1988}, which can be used to obtain high-probability lower bounds of the optimal value for a kind of optimization problem (For details, see~\cite{thrampoulidis2015d}). 
By using additional convexity assumptions on the objective function, CGMT enables us to obtain tighter results in the analysis.  
Specifically, CGMT associates the primary optimization (PO) problem with the auxiliary optimization (AO) problem given by 
\begin{align}
	\text{(PO): }& 
	\Phi(\bm{G}) 
	= 
	\min_{\bm{e} \in \mathcal{S}_{\txe}} \max_{\bm{u} \in \mathcal{S}_{\txu}} \curbra{\bm{u}^{\top} \bm{G} \bm{e} + \xi(\bm{e}, \bm{u})}, \label{eq:PO_CGMT} \\
	\text{(AO): }& 
	\phi(\bm{g}, \bm{h}) 
	= 
	\min_{\bm{e} \in \mathcal{S}_{\txe}} \max_{\bm{u} \in \mathcal{S}_{\txu}} \left\{ \norm{\bm{e}}_{2} \bm{g}^{\top} \bm{u} - \norm{\bm{u}}_{2} \bm{h}^{\top} \bm{e} \right. \notag \\
	&\hspace{50mm} 
	\left. + \xi(\bm{e}, \bm{u}) \right\},  \label{eq:AO_CGMT} 
\end{align}
respectively, where $\bm{G} \in \mathbb{R}^{M \times N}$, $\bm{g} \in \mathbb{R}^{M}$, $\bm{h} \in \mathbb{R}^{N}$, $\mathcal{S}_{\txe} \subset \mathbb{R}^{N}$, $\mathcal{S}_{\txu} \subset \mathbb{R}^{M}$, and $\xi (\cdot, \cdot): \mathbb{R}^{N} \times \mathbb{R}^{M} \to \mathbb{R}$. 
$\mathcal{S}_{\txe}$ and $\calS_{\txu}$ are assumed to be closed compact sets. 
$\xi (\cdot,\cdot)$ is a continuous convex-concave function on $\mathcal{S}_{\txe} \times \calS_{\txu}$, i.e., $\xi (\cdot, \bm{u}): \mathcal{S}_{\txe} \to \mathbb{R}$ is convex for fixed $\bm{u} \in \calS_{\txu}$ and $\xi (\bm{e}, \cdot): \calS_{\txu} \to \mathbb{R}$ is concave for fixed $\bm{e} \in \mathcal{S}_{\txe}$. 
Also, the elements of $\bm{G}$, $\bm{g}$, and $\bm{h}$ are i.i.d.\ standard Gaussian random variables. 
From the following theorem, we can relate the optimizer $\hat{\bm{e}}_{\Phi}(\bm{G})$ of (PO) with the optimal value of (AO) in the large system limit of $M, N \to \infty$ with a fixed ratio $\Delta=M/N$. 
\begin{theorem}[CGMT~\cite{thrampoulidis2018a}] \label{th:CGMT} 
	Let $\calT$ be an open set in $\mathcal{S}_{\txe}$ and $\calT^{\txc} = \mathcal{S}_{\txe} \setminus \calS$. 
	Also, let $\phi_{\calT^{\txc}} (\bm{g}, \bm{h})$ be the optimal value of (AO) with the constraint $\bm{e} \in \calT^{\txc}$. 
	If there are constants $\eta > 0$ and $\bar{\phi}$ satisfying (i) $\phi (\bm{g}, \bm{h}) \le \bar{\phi} + \eta$ and (ii) $\phi_{\calT^{\txc}} (\bm{g}, \bm{h}) \ge \bar{\phi} + 2 \eta$ with probability approaching $1$, then we have ${\displaystyle \lim_{N \to \infty}} \Pr \paren{\hat{\bm{e}}_{\Phi} \in \calT} = 1$. 
\end{theorem}
Intuitively, CGMT enables us to analyze (AO) instead of (PO). 
The conditions (i) and (ii) in the theorem mean that the optimal value of (AO) is achieved in $\calT$ with probability approaching $1$. 
The theorem says that, under the conditions, the optimizer of the original (PO) is also in $\calT$. 
Hence, if the optimization problem to be analyzed can be written as (PO), we can analyze the optimizer of the corresponding (AO) instead of that of (PO). 
Since the bilinear term $\bm{u}^{\top} \bm{G} \bm{e}$ in (PO) is divided as $\norm{\bm{e}}_{2} \bm{g}^{\top} \bm{u} - \norm{\bm{u}}_{2} \bm{h}^{\top} \bm{e}$ in (AO), (AO) is easier to analyze than (PO) in usual. 

In the CGMT-based analyses~\cite{thrampoulidis2018}, the original optimization problem (e.g.,~\eqref{eq:optimization}) is firstly rewritten into the form of (PO). 
The optimizer of the corresponding (AO) is then analyzed, and the conditions (i) and (ii) are verified for a properly defined set $\calS$. 
Finally, the asymptotic result for the optimizer of the original optimization problem is obtained by using CGMT. 

\section{Derivation of Claim~\ref{cl:main}} \label{app:derivation}
\begin{figure*}[t]
    \centering
    \includegraphics[width=160mm]{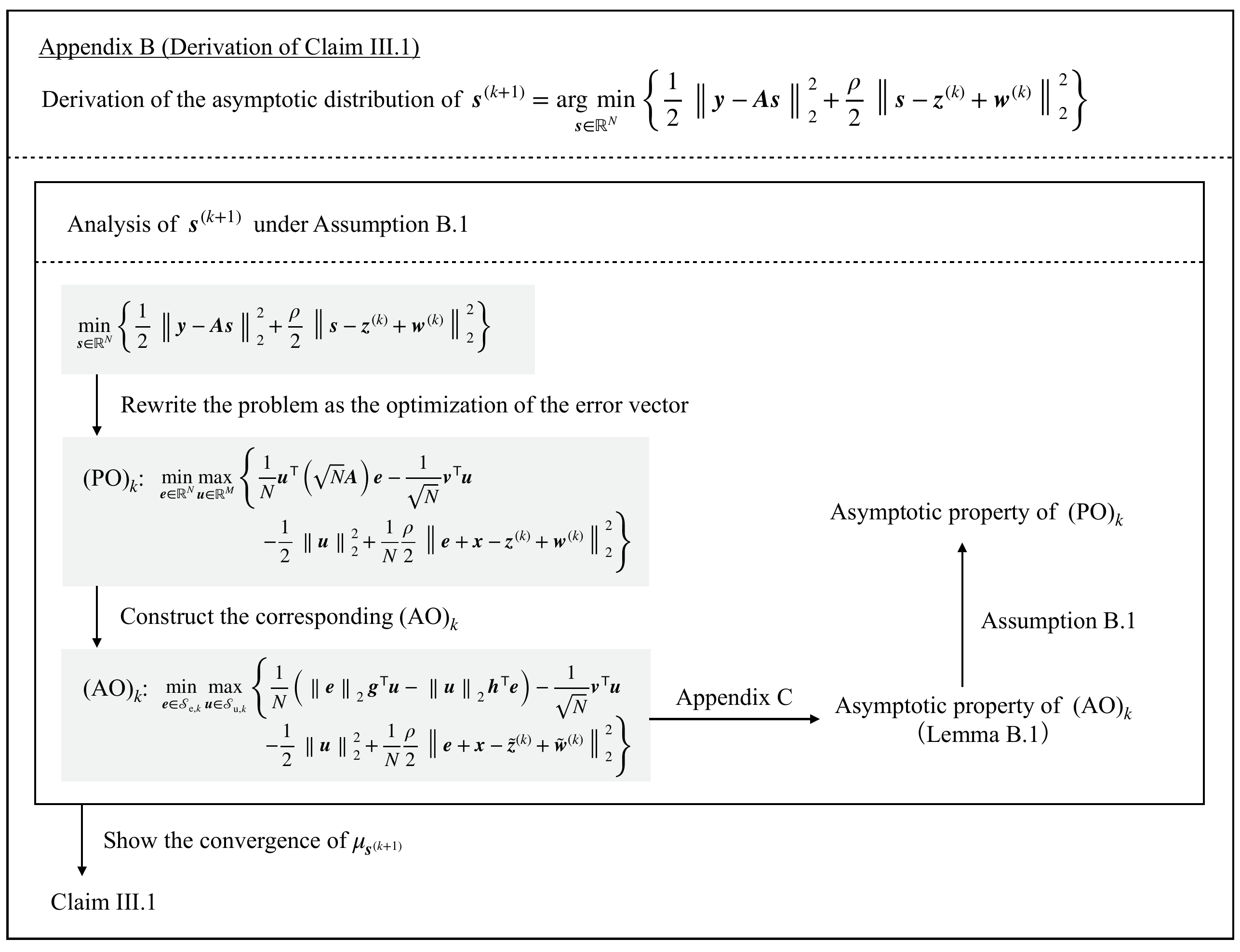}
    \caption{Overview of the derivation in Appendices.}
    \label{fig:derivation_overview}
\end{figure*}

In this section, we derive the result of Claim~\ref{cl:main}. 
The overview of the derivation is summarized in Fig.~\ref{fig:derivation_overview}. 
\subsection{Assumption}
For the derivation of Claim~\ref{cl:main}, we use the procedure of the CGMT-based analysis. 
However, the optimiztaion problem in~\eqref{eq:ADMM_update_s1} to be analyzed includes the regularization term with $\bm{z}^{(k)}$ and $\bm{w}^{(k)}$, which are not independent of the matrix $\bm{A}$. 
The standard CGMT-based analyses have been applied for the case with the regularization independent of $\bm{A}$, it would be required to generalize CGMT. 
Moreover, since we need to analyze the optimization problem in~\eqref{eq:ADMM_update_s1} iteratively, we assume the sufficient extension of CGMT for the derivation in this paper. 

We consider the sequence of the optimization problems given by 
\begin{align}
    \text{(PO)$_{k}$: }& 
    \Phi_{k+1}(\bm{G})
    = 
    \min_{\bm{e} \in \mathcal{S}_{\txe, k}} \max_{\bm{u} \in \mathcal{S}_{\txu, k}} \left\{ \bm{u}^{\top} \bm{G} \bm{e} \right. \notag \\
    &\hspace{25mm} 
    \left. + \xi_{k}(\bm{e}, \bm{u}, \hat{\bm{e}}_{\Phi}^{(1)}, \ldots, \hat{\bm{e}}_{\Phi}^{(k)}) \right\}, \label{eq:PO_k} \\
    \text{(AO)$_{k}$: }& 
    \phi_{k+1}(\bm{g}, \bm{h}) 
    = 
    \min_{\bm{e} \in \mathcal{S}_{\txe, k}} \max_{\bm{u} \in \mathcal{S}_{\txu, k}} \left\{ \norm{\bm{e}}_{2} \bm{g}^{\top} \bm{u} - \norm{\bm{u}}_{2} \bm{h}^{\top} \bm{e} \right. \notag \\
    &\hspace{25mm} 
    \left. + \xi_{k}(\bm{e}, \bm{u}, \hat{\bm{e}}_{\phi}^{(1)}, \ldots, \hat{\bm{e}}_{\phi}^{(k)}) \right\}, \label{eq:AO_k}
\end{align}
where $\hat{\bm{e}}_{\Phi}^{(k+1)}$ and $\hat{\bm{e}}_{\phi}^{(k+1)}$ denote the optimal value of $\bm{e}$ for (PO)$_{k}$ and (AO)$_{k}$, respectively ($k = 0, 1, 2, \dotsc$). 
The other variables are defined similarly as in~\eqref{eq:PO_CGMT} and~\eqref{eq:AO_CGMT}. 
$\mathcal{S}_{\txe, k} \subset \mathbb{R}^{N}$ and $\mathcal{S}_{\txu, k} \subset \mathbb{R}^{M}$ are closed compact sets. 
The continuous function $\xi_{k} (\cdot)$ is convex for $\bm{e}$ and concave for $\bm{u}$. 
We assume an iterative version of CGMT as follows. 
\begin{assumption}[iterative CGMT] \label{ass:iterativeCGMT} 
	For each $k = 0, 1, 2, \dotsc$, let $\calT_{k}$ be an open set in $\calS_{\txe, k}$ and $\calT_{k}^{\txc} = \calS_{\txe, k} \setminus \calT_{k}$. 
	Also, let $\phi_{k+1, \calT_{k}^{\txc}} (\bm{g}, \bm{h})$ be the optimal value of (AO)$_{k}$ with the constraint $\bm{e} \in \calT_{k}^{\txc}$. 
	If there are constants $\eta_{k} > 0$ and $\bar{\phi}_{k}$ satisfying (i) $\phi_{k+1} (\bm{g}, \bm{h}) \le \bar{\phi}_{k} + \eta_{k}$ and (ii) $\phi_{k+1, \calT_{k}^{\txc}} (\bm{g}, \bm{h}) \ge \bar{\phi}_{k} + 2 \eta_{k}$ with probability approaching $1$ for all $k \le K$, then we have ${\displaystyle \lim_{N \to \infty}} \Pr \paren{ \hat{\bm{e}}_{\Phi}^{(k+1)} \in \calT_{k} } = 1$ for any $k \le K$. 
\end{assumption}

In the assumption, the sequence $\Phi_{1}(\bm{G}), \Phi_{2}(\bm{G}), \dotsc$ shares the matrix $\bm{G}$ to handle the iterative update equations of ADMM. 
Similarly, the corresponding sequence $\phi_{1}(\bm{g}, \bm{h}), \phi_{2}(\bm{g}, \bm{h}), \dotsc$ shares the vectors $\bm{g}$ and $\bm{h}$. 
Moreover, we also allow that the function $\xi_{k}(\cdot)$ in (PO)$_{k}$ (resp. (AO)$_{k}$) includes $\hat{\bm{e}}_{\Phi}^{(1)}, \ldots, \hat{\bm{e}}_{\Phi}^{(k)}$ (resp. $\hat{\bm{e}}_{\phi}^{(1)}, \ldots, \hat{\bm{e}}_{\phi}^{(k)}$), which are the solutions of $\bm{e}$ at the previous iterations. 
If we consider only (PO)$_{0}$ and (AO)$_{0}$, the above assumption is equivalent to the case of the original CGMT, and hence the result is proven rigorously~\cite{thrampoulidis2018}. 

\subsection{Construction of (PO)$_{k}$ and (AO)$_{k}$}
To apply Assumption~\ref{ass:iterativeCGMT} to the derivation of Claim~\ref{cl:main}, we obtain the problems (PO)$_{k}$ and (AO)$_{k}$ corresponding to the update of $\bm{s}^{(k)}$ in~\eqref{eq:ADMM_update_s1}. 
We firstly define the error vector $\bm{e} = \bm{s}-\bm{x}$ to rewrite the optimization problem~\eqref{eq:ADMM_update_s1} as 
\begin{align}
    \min_{\bm{e} \in \mathbb{R}^{N}} 
    \frac{1}{N} 
    \curbra{
        \frac{1}{2} \norm{\bm{A}\bm{e}-\bm{v}}_{2}^{2} + \frac{\rho}{2} \norm{ \bm{e} + \bm{x} - \bm{z}^{(k)} + \bm{w}^{(k)} }_{2}^{2} 
    }, 
\end{align}
where the objective function is normalized by $N$. 
By using 
\begin{align}
    \frac{1}{2} \norm{\bm{A} \bm{e} - \bm{v}}_{2}^{2} 
    &= 
    \max_{\bm{u} \in \mathbb{R}^{M}} 
    \curbra{
        \sqrt{N} \bm{u}^{\top} (\bm{A} \bm{e} - \bm{v}) - \frac{N}{2} \norm{\bm{u}}_{2}^{2} 
    }, 
\end{align}
we can obtain the equivalent (PO)$_{k}$ problem given by 
\begin{align}
    &\min_{\bm{e} \in \mathbb{R}^{N}} \max_{\bm{u} \in \mathbb{R}^{M}} 
    \Biggl\{
        \frac{1}{N} \bm{u}^{\top}(\sqrt{N} \bm{A}) \bm{e} 
        - \frac{1}{\sqrt{N}} \bm{v}^{\top}\bm{u} - \frac{1}{2} \norm{\bm{u}}_{2}^{2} \notag \\\
    &\hspace{26mm}
        + \frac{1}{N} \frac{\rho}{2} \norm{ \bm{e} + \bm{x} - \bm{z}^{(k)} + \bm{w}^{(k)} }_{2}^{2} 
    \Biggr\}. \label{eq:PO_k_ADMM}
\end{align}

Note that the optimal value $\hat{\bm{e}}_{\Phi}^{(k+1)}$ of $\bm{e}$ in~\eqref{eq:PO_k_ADMM} can be written as $\hat{\bm{e}}_{\Phi}^{(k+1)} = \bm{s}^{(k+1)} - \bm{x}$ from the definition of $\bm{e}$ and~\eqref{eq:ADMM_update_s1}. 
Since we have 
\begin{align}
    \bm{z}^{(k+1)} 
    &= 
    \prox_{\frac{\lambda}{\rho} f} \paren{\hat{\bm{e}}_{\Phi}^{(k+1)} + \bm{x} + \bm{w}^{(k)}}, \label{eq:z_update_by_e} \\
    \bm{w}^{(k+1)} 
    &= 
    \bm{w}^{(k)} + \hat{\bm{e}}_{\Phi}^{(k+1)} + \bm{x} - \bm{z}^{(k+1)}, \label{eq:w_update_by_e}
\end{align}
from~\eqref{eq:ADMM_update_z2} and~\eqref{eq:ADMM_update_w}, $\bm{z}^{(k)}$ and $\bm{w}^{(k)}$ can be written with $\hat{\bm{e}}_{\Phi}^{(1)}, \dotsc, \hat{\bm{e}}_{\Phi}^{(k)}$ by using~\eqref{eq:z_update_by_e} and~\eqref{eq:w_update_by_e} iteratively ($\bm{z}^{(0)} = \bm{w}^{(0)} = \bm{0}$). 
We can thus see that the optimization problem~\eqref{eq:PO_k_ADMM} has the form of (PO)$_{k}$ in~\eqref{eq:PO_k}\footnote{
Although the constraint set of the problem~\eqref{eq:PO_k_ADMM} is unbounded, we can introduce a bounded constraint with sufficiently large convex constraint sets $\calS_{\txe, k}$ and $\calS_{\txu, k}$ to apply CGMT (For details, see~\cite[Appendix A]{thrampoulidis2018}). }. 

The corresponding (AO)$_{k}$ problem is given by 
\begin{align}
    &\min_{\bm{e} \in \calS_{\txe, k}} \max_{\bm{u} \in \calS_{\txu, k}}
    \Biggl\{
        \frac{1}{N} \paren{\norm{\bm{e}}_{2} \bm{g}^{\top} \bm{u} - \norm{\bm{u}}_{2} \bm{h}^{\top} \bm{e}} 
        - \frac{1}{\sqrt{N}} \bm{v}^{\top}\bm{u}  \notag \\\
    &\hspace{10mm}
        - \frac{1}{2} \norm{\bm{u}}_{2}^{2}
        + \frac{1}{N} \frac{\rho}{2} \norm{ \bm{e} + \bm{x} - \tilde{\bm{z}}^{(k)} + \tilde{\bm{w}}^{(k)} }_{2}^{2} 
    \Biggr\}, \label{eq:AO_k_ADMM}
\end{align}
where $\tilde{\bm{z}}^{(k)}$ and $\tilde{\bm{w}}^{(k)}$, i.e., the counterparts of $\bm{z}^{(k)}$ and $\bm{w}^{(k)}$ in~\eqref{eq:PO_k_ADMM}, are given by the recursion 
\begin{align}
    \tilde{\bm{z}}^{(k+1)} 
    &= 
    \prox_{\frac{\lambda}{\rho} f} \paren{\hat{\bm{e}}_{\phi}^{(k+1)} + \bm{x} + \tilde{\bm{w}}^{(k)}}, \label{eq:z_update_by_e_AO} \\
    \tilde{\bm{w}}^{(k+1)} 
    &= 
    \tilde{\bm{w}}^{(k)} + \hat{\bm{e}}_{\phi}^{(k+1)} + \bm{x} - \tilde{\bm{z}}^{(k+1)} \label{eq:w_update_by_e_AO}
\end{align}
with $\tilde{\bm{z}}^{(0)} = \tilde{\bm{w}}^{(0)} = \bm{0}$. 
The update of $\hat{\bm{e}}_{\phi}^{(k+1)}$ will be discussed later in~\eqref{eq:e_hat_AO}. 

\subsection{Analytical Result for (AO)$_{k}$}
For the (AO)$_{k}$ problem above, we have the following lemma. 
For the proof of the lemma, we do not require Assumption~\ref{ass:iterativeCGMT}. 
\begin{lemma} \label{lem:analysis_AO_k}
    Let 
    \begin{align}
        \mathcal{L} 
        &= 
        \{ 
            \psi(\cdot, \cdot): \mathbb{R} \times \mathbb{R} \to \mathbb{R} 
            \mid \notag\\
        &\hspace{7mm}
            \psi(\cdot, x) \text{ is Lipschitz continuous for any $x \in \mathbb{R}$}
        \} \label{eq:set_Lipschitz}
    \end{align}
    and 
    \begin{align}
        \calT_{k}
        &=
        \Biggl\{ 
            \bm{t} \in \mathbb{R}^{N} 
            \ \Bigg|\ \notag \\
        &\hspace{3mm}
            \abs{ 
                \frac{1}{N} \sum_{n=1}^{N} \psi(t_{n}, x_{n}) 
                - \Ex{\psi\paren{S_{k+1} - X, X}}
            } 
            < \varepsilon 
        \Biggr\}. \label{eq:set_Tk}
    \end{align}
    For any fixed non-negative integer $K$ and $\psi(\cdot, \cdot) \in \mathcal{L}$, the (AO)$_{k}$ problem in~\eqref{eq:AO_k_ADMM} satisfies the conditions (i) and (ii) in Assumption~\ref{ass:iterativeCGMT} with $\calT_{k}$ in~\eqref{eq:set_Tk} for $k = 0, 1, \dotsc, K$. 
\end{lemma}
\begin{IEEEproof}
See Appendix~\ref{app:proof_lemma}. 
\end{IEEEproof}

\subsection{Convergence of $\mu_{\bm{s}^{(k+1)}}$}
From Assumption~\ref{ass:iterativeCGMT} and Lemma~\ref{lem:analysis_AO_k}, we can obtain ${\displaystyle \lim_{N \to \infty}} \Pr \paren{ \hat{\bm{e}}_{\Phi}^{(k+1)} \in \calT_{k} } = 1$. 
Since $\hat{\bm{e}}_{\Phi}^{(k+1)} = \bm{s}^{(k+1)} - \bm{x}$, we have 
\begin{align}
    &\plim_{N \to \infty}
    \frac{1}{N} \sum_{n=1}^{N} \psi\paren{s_{n}^{(k+1)} - x_{n}, x_{n}} \notag \\
    &= 
    \Ex{\psi\paren{S_{k+1} - X, X}}. \label{eq:convergence_psi}
\end{align}

Finally, we show 
\begin{align}
    \lim_{N \to \infty} 
    \Pr 
    \paren{
        \abs{
            \int g d \mu_{\bm{s}^{(k+1)}} - \int g d \mu_{S_{k+1}} 
        }
        < 
        \varepsilon
    }
    = 
    1 \label{eq:goal_proof}
\end{align}
for any continuous compactly supported function $g(\cdot): \mathbb{R} \to \mathbb{R}$ and any $\varepsilon$ ($>0$). 
Since the function $g(\cdot)$ has a compact support, there exists a polynomial $\nu(\cdot): \mathbb{R} \to \mathbb{R}$ such that 
\begin{align}
    \abs{g(x) - \nu(x)} < \frac{\varepsilon}{3}
\end{align}
for any $x$ in the support from the Stone-Weierstrass theorem~\cite{perez2008}. 
We thus have 
\begin{align}
    &\abs{
        \int g d \mu_{\bm{s}^{(k+1)}} - \int g d \mu_{S_{k+1}} 
    } \notag \\
    &< 
    \abs{
        \int g d \mu_{\bm{s}^{(k+1)}} - \int \nu d \mu_{\bm{s}^{(k+1)}} 
    } \notag \\
    &\hspace{4mm}
    + 
    \abs{
        \int \nu d \mu_{\bm{s}^{(k+1)}} - \int \nu d \mu_{S_{k+1}} 
    } \notag \\
    &\hspace{4mm}+ 
    \abs{
        \int \nu d \mu_{S_{k+1}} - \int g d \mu_{S_{k+1}} 
    } \\
    &< 
    \abs{
        \int \nu d \mu_{\bm{s}^{(k+1)}} - \int \nu d \mu_{S_{k+1}} 
    } 
    + 
    \frac{2}{3} \varepsilon. \label{eq:diff_integral}
\end{align}
Since the polynomial $\nu(\cdot)$ is Lipschitz on the compact support, we define $\psi(e, x) = \nu(e + x)$ in~\eqref{eq:convergence_psi} and obtain 
\begin{align}
    \plim_{N \to \infty} 
    \frac{1}{N} \sum_{n=1}^{N} \nu\paren{s_{n}^{(k+1)}} 
    &= 
    \Ex{\nu\paren{S_{k+1}}}. \label{eq:convergence_nu}
\end{align}
Since we have~\eqref{eq:goal_proof} from~\eqref{eq:diff_integral} and~\eqref{eq:convergence_nu}, we can obtain $\int g d \mu_{\bm{s}^{(k+1)}} \Pto \int g d \mu_{S_{k+1}}$ as $N \to \infty$, which is the result of Claim~\ref{cl:main}. 
\section{Proof of Lemma~\ref{lem:analysis_AO_k}} \label{app:proof_lemma}
We here prove Lemma~\ref{lem:analysis_AO_k}. 
Since the analysis for the optimization problem~\eqref{eq:AO_k_ADMM} follows the standard approach with CGMT~\cite{thrampoulidis2018}, we omit some details and show only the outline of the proof in some parts. 
For details of the CGMT-based analysis, see~\cite{thrampoulidis2018,thrampoulidis2018a,atitallah2017,hayakawa2020a} and references therein. 
\subsection{Analysis of (AO)$_{k}$ Problem}
We analyze the (AO)$_{k}$ problem in~\eqref{eq:AO_k_ADMM}. 
Since both $\bm{g}$ and $\bm{v}$ are Gaussian, the vector $\frac{\norm{\bm{e}}_{2}}{\sqrt{N}}\bm{g} - \bm{v}$ is also Gaussian with zero mean and the covariance matrix $\paren{\frac{\norm{\bm{e}}_{2}^{2}}{N} + \sigma_{\txv}^{2}} \bm{I}$. 
Hence, we can rewrite $\paren{\frac{\norm{\bm{e}}_{2}}{\sqrt{N}}\bm{g}-\bm{v}}^{\top} \bm{u}$ as  $\sqrt{\frac{\norm{\bm{e}}_{2}^{2}}{N} + \sigma_{\txv}^{2}} \bm{g}^{\top} \bm{u}$ with the slight abuse of notation, where $\bm{g}$ has i.i.d.\ standard Gaussian elements. 
We apply this technique to~\eqref{eq:AO_k_ADMM} and obtain
\begin{align}
    &\min_{\bm{e} \in \calS_{\txe, k}} \max_{\bm{u} \in \calS_{\txu, k}}
    \Biggl\{
        \frac{1}{\sqrt{N}} \sqrt{\frac{\norm{\bm{e}}_{2}^{2}}{N} + \sigma_{\txv}^{2}} \bm{g}^{\top} \bm{u} 
        - \frac{1}{N} \norm{\bm{u}}_{2} \bm{h}^{\top} \bm{e} \notag \\
    &\hspace{10mm}
        - \frac{1}{2} \norm{\bm{u}}_{2}^{2}
        + \frac{1}{N} \frac{\rho}{2} \norm{ \bm{e} + \bm{x} - \tilde{\bm{z}}^{(k)} + \tilde{\bm{w}}^{(k)}}_{2}^{2} 
    \Biggr\}. \label{eq:AO_supplement1}
\end{align}
When we denote $\norm{\bm{u}}_{2} = \beta$, the maximum value of $\bm{g}^{\top} \bm{u}$ in the first term is given by $\beta \norm{\bm{g}}_{2}$. 
We can thus rewrite~\eqref{eq:AO_supplement1} as 
\begin{align}
    &\min_{\bm{e} \in \calS_{\txe, k}} \max_{0 \le \beta \le C_{\txu, k}}
    \Biggl\{
        \sqrt{\frac{\norm{\bm{e}}_{2}^{2}}{N} + \sigma_{\txv}^{2}} \frac{\beta \norm{\bm{g}}_{2}}{\sqrt{N}}  
        - \frac{\beta}{N} \bm{h}^{\top} \bm{e} \notag \\
    &\hspace{10mm}
        - \frac{1}{2} \beta^{2}
        + \frac{1}{N} \frac{\rho}{2} \norm{ \bm{e} + \bm{x} - \tilde{\bm{z}}^{(k)} + \tilde{\bm{w}}^{(k)}}_{2}^{2} 
    \Biggr\}, \label{eq:AO_supplement2}
\end{align}
where $C_{\txu, k}$ is a sufficiently large constant satisfying $C_{\txu, k} \ge \max_{\bm{u} \in \calS_{\txu, k}} \norm{\bm{u}}_{2}$. 
We then use the identity 
\begin{align}
    \chi 
    &= 
    \min_{\alpha>0}
    \paren{
        \frac{\alpha}{2} + \frac{\chi^{2}}{2\alpha} 
    } 
\end{align}
for $\chi$ ($>0$) to transform the square root term in~\eqref{eq:AO_supplement2} as 
\begin{align}
    \sqrt{\frac{\norm{\bm{e}}_{2}^{2}}{N} + \sigma_{\txv}^{2}}
    &= 
    \min_{\alpha>0}
    \paren{
        \frac{\alpha}{2} + \frac{\frac{\norm{\bm{e}}_{2}^{2}}{N} + \sigma_{\txv}^{2}}{2\alpha} 
    }. \label{eq:root_eliminate_trick}
\end{align}
By substituting~\eqref{eq:root_eliminate_trick} into~\eqref{eq:AO_supplement2}, we have 
\begin{align}
    &\hspace{-4mm}\min_{\bm{e} \in \calS_{\txe, k}} \max_{0 \le \beta \le C_{\txu, k}} \min_{\alpha>0} 
    \Biggl\{
        \frac{\alpha\beta}{2} \frac{\norm{\bm{g}}_{2}}{\sqrt{N}} 
        + \frac{1}{N} \frac{\beta}{2 \alpha} \frac{\norm{\bm{g}}_{2}}{\sqrt{N}} \norm{\bm{e}}_{2}^{2}
        + \frac{\beta\sigma_{\txv}^{2}}{2\alpha} \frac{\norm{\bm{g}}_{2}}{\sqrt{N}} \notag \\
    &\hspace{2mm}
        - \frac{\beta}{N} \bm{h}^{\top} \bm{e}
        - \frac{1}{2} \beta^{2}
        + \frac{1}{N} \frac{\rho}{2} \norm{ \bm{e} + \bm{x} - \tilde{\bm{z}}^{(k)} + \tilde{\bm{w}}^{(k)} }_{2}^{2} 
    \Biggr\}. \label{eq:AO_supplement3}
\end{align}
We can further rewrite~\eqref{eq:AO_supplement3} as\footnote{
    For the optimization problem in~\eqref{eq:AO_supplement3}, we have 
    \begin{itemize}
        \item The objective function is convex for $\alpha, \bm{e}$ and concave for $\beta$. 
        \item The constraint sets for $\beta$ and $\bm{e}$ are convex compact sets. 
        \item Since the optimal value of $\alpha$ in~\eqref{eq:root_eliminate_trick} can be written as $\hat{\alpha} = \sqrt{\frac{\norm{\bm{e}}_{2}^{2}}{N} + \sigma_{\txv}^{2}} > \sigma_{\txv}$, the result of the optimization is the same if we change the range of $\alpha$ into the compact set $\alpha \in [\sigma_{\txv}, C]$, where $C$ is a sufficiently large constant. 
    \end{itemize}
    Thus, from the minimax theorem, we exchange the order of the minimization and the maximization in~\eqref{eq:AO2}. 
    Moreover, we change the range of $\beta$ from $[0, C_{\txu, k}]$ to $[0, \infty)$ without changing the result of the optimization. 
}
\begin{align}
    &\min_{\alpha>0} \max_{\beta \ge 0} 
    \Biggl\{
        \frac{\alpha\beta}{2} \frac{\norm{\bm{g}}_{2}}{\sqrt{N}} 
        + \frac{\beta\sigma_{\txv}^{2}}{2\alpha} \frac{\norm{\bm{g}}_{2}}{\sqrt{N}}
        - \frac{1}{2} \beta^{2} \notag \\
    &\hspace{30mm}
        + 
        \min_{\bm{e} \in \calS_{\txe, k}} 
        \frac{1}{N} \sum_{n=1}^{N} J_{n}^{(k+1)}(e_{n}, \alpha, \beta) \label{eq:AO2}
    \Biggr\}, 
\end{align}
where the subscript $(\cdot)_{n}$ denotes the $n$-th element of the corresponding bold vector and 
\begin{align}
    J_{n}^{(k+1)} (e_{n}, \alpha, \beta) 
    &= 
    \frac{\beta}{2\alpha} \frac{\norm{\bm{g}}_{2}}{\sqrt{N}} e_{n}^{2} 
    - \beta h_{n} e_{n} \notag \\
    &\hspace{5mm}
    + \frac{\rho}{2} \paren{ e_{n} + x_{n} - \tilde{z}_{n}^{(k)} + \tilde{w}_{n}^{(k)} }^{2}.
\end{align}
The minimum value of $J_{n}^{(k+1)}(e_{n}, \alpha, \beta)$ is achieved when 
\begin{align}
    \hat{e}_{n}^{(k+1)} (\alpha, \beta) 
    &= 
    \dfrac{1}{\dfrac{\beta}{\alpha}\dfrac{\norm{\bm{g}}_{2}}{\sqrt{N}} + \rho} 
    \paren{\beta h_{n} - \rho \paren{x_{n} - \tilde{z}_{n}^{(k)} + \tilde{w}_{n}^{(k)}} }. \label{eq:e_hat}
\end{align}
We then define $\hat{s}_{n}^{(k+1)} (\alpha, \beta) = \hat{e}_{n}^{(k+1)} (\alpha, \beta) + x_{n}$, which is given by 
\begin{align}
    \hat{s}_{n}^{(k+1)} (\alpha, \beta) 
    &= 
    \frac{1}{\dfrac{\beta}{\alpha}\dfrac{\norm{\bm{g}}_{2}}{\sqrt{N}} + \rho} 
    \Biggl(
        \frac{\beta}{\alpha}\frac{\norm{\bm{g}}_{2}}{\sqrt{N}} \paren{x_{n} +  \frac{\sqrt{N}}{\norm{\bm{g}}_{2}} \alpha h_{n}} \notag \\
    &\hspace{30mm}
    + \rho \paren{\tilde{z}_{n}^{(k)} - \tilde{w}_{n}^{(k)}}
    \Biggr). \label{eq:s_hat}
\end{align}
The optimization problem~\eqref{eq:AO2} can be rewritten as 
\begin{align}
    &\min_{\alpha>0} \max_{\beta \ge 0} 
    \Biggl\{
        \frac{\alpha\beta}{2} \frac{\norm{\bm{g}}_{2}}{\sqrt{N}} 
        + \frac{\beta\sigma_{\txv}^{2}}{2\alpha} \frac{\norm{\bm{g}}_{2}}{\sqrt{N}}
        - \frac{1}{2} \beta^{2} \notag \\
    &\hspace{12mm}
        + 
        \frac{1}{N} \sum_{n=1}^{N} J_{n}^{(k+1)} \paren{\hat{s}_{n}^{(k+1)} (\alpha, \beta)-x_{n}, \alpha, \beta} \label{eq:AO3}
    \Biggr\}. 
\end{align}
We denote the optimal value of the objective function in~\eqref{eq:AO3} and the corresponding solution as $\phi_{k, N}^{\ast}$ and $\paren{\alpha_{k, N}^{\ast}, \beta_{k, N}^{\ast}}$, respectively. 
The optimal value of $\bm{e}$ in (AO)$_{k}$ can be written as 
\begin{align}
    \hat{\bm{e}}_{\phi}^{(k+1)} 
    &= 
    \sqbra{\hat{e}_{1}^{(k+1)}(\alpha_{k, N}^{\ast}, \beta_{k, N}^{\ast})\ \dotsb\ \hat{e}_{N}^{(k+1)}(\alpha_{k, N}^{\ast}, \beta_{k, N}^{\ast})}^{\top} \\
    &= 
    \dfrac{1}{\dfrac{\beta_{k, N}^{\ast}}{\alpha_{k, N}^{\ast}}\dfrac{\norm{\bm{g}}_{2}}{\sqrt{N}} + \rho} 
    \paren{\beta_{k, N}^{\ast} \bm{h} - \rho \paren{\bm{x} - \tilde{\bm{z}}^{(k)} + \tilde{\bm{w}}^{(k)}} }. \label{eq:e_hat_AO}
\end{align}

We then show the objective function of~\eqref{eq:AO3} converges pointwise to 
\begin{align}
    \frac{\alpha\beta\sqrt{\Delta}}{2} 
    + \frac{\beta \sigma_{\txv}^{2} \sqrt{\Delta}}{2\alpha} 
    - \frac{1}{2} \beta^{2} 
    + \Ex{J^{(k+1)}(\alpha,\beta)}, \label{eq:AO_converge}
\end{align}
where 
\begin{align}
    &J^{(k+1)} (\alpha, \beta) \notag \\
    &= 
    \frac{\beta\sqrt{\Delta}}{2\alpha} \paren{\hat{S}_{k+1}(\alpha, \beta) - X}^{2} 
    - \beta H \paren{\hat{S}_{k+1}(\alpha, \beta) - X} \notag \\
    &\hspace{4mm}+ 
    \frac{\rho}{2} \paren{ \hat{S}_{k+1}(\alpha, \beta) - Z_{k} + W_{k} }^{2} 
\end{align}
and $\hat{S}_{k+1}(\alpha, \beta)$ is defined in~\eqref{eq:S_alpha_beta}. 
Note that the function~\eqref{eq:AO_converge} is the objective function of~\eqref{eq:SO} in Claim~\ref{cl:main}. 
First, for $k=0$, we have 
\begin{align}
    \hat{s}_{n}^{(1)} (\alpha, \beta) 
    &= 
    \frac{\dfrac{\beta}{\alpha}\dfrac{\norm{\bm{g}}_{2}}{\sqrt{N}}}{\dfrac{\beta}{\alpha}\dfrac{\norm{\bm{g}}_{2}}{\sqrt{N}} + \rho} 
    \paren{x_{n} +  \frac{\sqrt{N}}{\norm{\bm{g}}_{2}} \alpha h_{n}} \label{eq:s_hat_1}
\end{align}
because $\tilde{z}_{n}^{(0)} = \tilde{w}_{n}^{(0)} = 0$. 
Since $\norm{\bm{g}}_{2} / \sqrt{N} \to \sqrt{\Delta}$ as $N \to \infty$, $\hat{s}_{n}^{(1)} (\alpha, \beta)$ can be modeled as the random variable $\hat{S}_{1}(\alpha, \beta)$ in the asymptotic regime. 
This means that $\frac{1}{N} \sum_{n=1}^{N} J_{n}^{(1)} \paren{\hat{s}_{n}^{(1)} (\alpha, \beta)-x_{n}, \alpha, \beta}$ converges to $\Ex{J^{(1)} (\alpha, \beta)}$ as $N \to \infty$ ($Z_{0}=W_{0}=0$), and hence the objective function in~\eqref{eq:AO3} converges to~\eqref{eq:AO_converge} when $k = 0$. 
We then consider the case with $k > 0$. 
By a similar discussion to~\cite[Lemma IV.1]{thrampoulidis2018a}, we have $\phi_{k, N}^{\ast} \Pto \phi_{k}^{\ast}$ and $(\alpha_{k,N}^{\ast}, \beta_{k,N}^{\ast}) \Pto (\alpha_{k}^{\ast}, \beta_{k}^{\ast})$ as $N \to \infty$, where $\phi_{k}^{\ast}$ and $(\alpha_{k}^{\ast}, \beta_{k}^{\ast})$ are the optimal values of the objective function and $(\alpha, \beta)$ in~\eqref{eq:SO}, respectively. 
Thus, $\hat{s}_{n}^{(1)} (\alpha_{0}^{\ast}, \beta_{0}^{\ast})$ ($= \hat{e}_{n}^{(1)} (\alpha_{0}^{\ast}, \beta_{0}^{\ast}) + x_{n} = \hat{e}_{\phi, n}^{(1)} + x_{n}$) can be modeled as $S_{1} = \hat{S}_{1}(\alpha_{0}^{\ast}, \beta_{0}^{\ast})$ in the asymptotic regime. 
Since the updates in~\eqref{eq:z_update_by_e_AO} and~\eqref{eq:w_update_by_e_AO} are element-wise and same as~\eqref{eq:update_Z},~\eqref{eq:update_W}, $\tilde{z}_{n}^{(1)}$ and $\tilde{w}_{n}^{(1)}$ can be modeled as the random variables $Z_{1}$ and $W_{1}$, respectively. 
By using such discussion iteratively for $k = 0, 1, \dotsc$, we can see that $\tilde{z}_{n}^{(k)}$ and $\tilde{w}_{n}^{(k)}$ can be modeled as the random variables $Z_{k}$ and $W_{k}$ also for $k > 0$, respectively. 
Thus, $\frac{1}{N} \sum_{n=1}^{N} J_{n}^{(k+1)} \paren{\hat{s}_{n}^{(k+1)} (\alpha, \beta)-x_{n}, \alpha, \beta}$ converges to $\Ex{J^{(k+1)} (\alpha, \beta)}$, and hence the objective function of~\eqref{eq:AO3} converges pointwise to the function~\eqref{eq:AO_converge}. 
\subsection{Conditions (i) and (ii) for (AO)$_{k}$}
We then consider the condition (i) in Assumption~\ref{ass:iterativeCGMT} for (AO)$_{k}$. 
As discussed in the previous paragraph, we have $\phi_{k, N}^{\ast} \Pto \phi_{k}^{\ast}$ as $N \to \infty$, where $\phi_{k, N}^{\ast}$ is the optimal value of the objective function in (AO)$_{k}$. 
Thus, the optimal value of (AO)$_{k}$ satisfies the condition (i) in Assumption~\ref{ass:iterativeCGMT} with probability approaching 1 for $\bar{\phi}_{k} = \phi_{k}^{\ast}$ and any $\eta_{k}$ ($>0$). 

Next, we investigate condition (ii) in Assumption~\ref{ass:iterativeCGMT}. 
Note again that the optimal value of $\bm{e}$ in (AO)$_{k}$ is given by $\hat{\bm{e}}_{\phi}^{(k+1)} = \sqbra{\hat{e}_{1}^{(k+1)}(\alpha_{k, N}^{\ast}, \beta_{k, N}^{\ast})\ \dotsb\ \hat{e}_{N}^{(k+1)}(\alpha_{k, N}^{\ast}, \beta_{k, N}^{\ast})}^{\top}$ from~\eqref{eq:AO2}--\eqref{eq:e_hat}. 
We use the following lemma to construct the set $\calT_{k}$ in~\eqref{eq:set_Tk}. 
\begin{lemma} \label{lem:exp}
    For any function $\psi(\cdot, \cdot) \in \mathcal{L}$ ($\mathcal{L}$ is defined in~\eqref{eq:set_Lipschitz}), we have 
    \begin{align}
        &\plim_{N \to \infty}
        \frac{1}{N} \sum_{n=1}^{N} \psi\paren{\hat{s}_{n}^{(k+1)}\paren{\alpha_{k,N}^{\ast}, \beta_{k,N}^{\ast}} - x_{n}, x_{n}} \notag \\
        &\hspace{10mm}= 
        \Ex{\psi\paren{\hat{S}_{k+1}\paren{\alpha_{k}^{\ast}, \beta_{k}^{\ast}} - X, X}}
    \end{align}
\end{lemma}
\begin{IEEEproof}
See Appendix~\ref{app:exp}. 
\end{IEEEproof}

From Lemma~\ref{lem:exp} and~\eqref{eq:set_Tk}, 
we can obtain $\hat{\bm{e}}_{\phi}^{(k+1)} = \hat{\bm{s}}^{(k+1)}(\alpha_{k, N}^{\ast}, \beta_{k, N}^{\ast}) - \bm{x} \in \calT_{k}$ with probability approaching $1$ for any $\varepsilon$ ($>0$). 
By using the strong convexity of $J_{n}^{(k+1)}(e_{n}, \alpha, \beta)$ over $e_{n}$, we can see that there exists a constant $\eta_{k}$ satisfying the condition (ii) in Assumption~\ref{ass:iterativeCGMT} with $\calT_{k}$. 
\section{Proof of Lemma~\ref{lem:exp}} \label{app:exp}
Define 
\begin{align}
    &\bar{s}_{n}^{(k+1)} (\alpha, \beta) \notag \\
    &= 
    \dfrac{1}{\dfrac{\beta\sqrt{\Delta}}{\alpha} + \rho} 
    \paren{ 
        \dfrac{\beta \sqrt{\Delta}}{\alpha} 
        \paren{x_{n} + \dfrac{\alpha}{\sqrt{\Delta}} h_{n}}
        + \rho \paren{z_{n}^{(k)} - w_{n}^{(k)}} 
    }, 
\end{align}
where we replace $\dfrac{\norm{\bm{g}}_{2}}{\sqrt{N}}$ in~\eqref{eq:s_hat} with its asymptotic value $\sqrt{\Delta}$. 
From the law of large numbers, we have 
\begin{align}
    &\plim_{N \to \infty}
    \frac{1}{N} \sum_{n=1}^{N} \psi\paren{\bar{s}_{n}^{(k+1)}\paren{\alpha_{k}^{\ast}, \beta_{k}^{\ast}} - x_{n}, x_{n}} \notag \\
    &\hspace{10mm}= 
    \Ex{\psi\paren{\hat{S}_{k+1}\paren{\alpha_{k}^{\ast}, \beta_{k}^{\ast}} - X, X}}. 
\end{align}
Thus, it is sufficient to show 
\begin{align}
    &\plim_{N \to \infty}
    \Biggl| 
        \frac{1}{N} \sum_{n=1}^{N} \psi\paren{\hat{s}_{n}^{(k+1)}\paren{\alpha_{k, N}^{\ast}, \beta_{k, N}^{\ast}} - x_{n}, x_{n}} \notag \\
    &\hspace{10mm}
        - 
        \frac{1}{N} \sum_{n=1}^{N} \psi\paren{\bar{s}_{n}^{(k+1)}\paren{\alpha_{k}^{\ast}, \beta_{k}^{\ast}} - x_{n}, x_{n}}
    \Biggr|
    = 
    0. 
\end{align}
Since $\psi(\cdot, x_{n})$ is Lipschitz, there is a constant $C_{\psi}$ ($>0$) such that 
\begin{align}
    &\Biggl| 
        \frac{1}{N} \sum_{n=1}^{N} \psi\paren{\hat{s}_{n}^{(k+1)}\paren{\alpha_{k, N}^{\ast}, \beta_{k, N}^{\ast}} - x_{n}, x_{n}} \notag \\
    &\hspace{5mm}
        - 
        \frac{1}{N} \sum_{n=1}^{N} \psi\paren{\bar{s}_{n}^{(k+1)}\paren{\alpha_{k}^{\ast}, \beta_{k}^{\ast}} - x_{n}, x_{n}}
    \Biggr| \notag \\
    &\leq 
    \frac{C_{\psi}}{N} \sum_{n=1}^{N} 
    \abs{
        \hat{s}_{n}^{(k+1)}\paren{\alpha_{k, N}^{\ast}, \beta_{k, N}^{\ast}} 
        - \bar{s}_{n}^{(k+1)}\paren{\alpha_{k}^{\ast}, \beta_{k}^{\ast}}
    } \\
    &\Pto 
    0\quad (N \to \infty), 
\end{align}
which completes the proof. 
\ifCLASSOPTIONcaptionsoff
  \newpage
\fi
%
\bibliographystyle{IEEEtran}
\bibliography{mybib}
%
%
\begin{IEEEbiography}{Ryo Hayakawa}
received the bachelor's degree in engineering, the master's degree in informatics, and Ph.D.\ degree in informatics from Kyoto University, Kyoto, Japan, in 2015, 2017, and 2020, respectively. 
He is currently an Assistant Professor at Graduate School of Engineering Science, Osaka University. 
He was a Research Fellow (DC1) of the Japan Society for the Promotion of Science (JSPS) from 2017 to 2020. 
He received the 33rd Telecom System Technology Student Award, APSIPA ASC 2019 Best Special Session Paper Nomination Award, and the 16th IEEE Kansai Section Student Paper Award. 
His research interests include signal processing and mathematical optimization. 
He is a member of IEICE. 
\end{IEEEbiography}
\vfill
\end{document}